\documentclass[a4paper,11pt]{article}
\usepackage[latin1]{inputenc}
\usepackage[T1]{fontenc}
\usepackage[english]{babel}
\usepackage[hmargin={32mm,32mm},vmargin={32mm,35mm}]{geometry}
\usepackage{amsmath}
\usepackage{amsfonts}
\usepackage{amssymb}
\usepackage{enumerate}
\usepackage{graphicx}
\usepackage{xypic}
\usepackage{tocloft}
\usepackage{bbm}
\usepackage{feynmf}
\usepackage{epsfig}
\usepackage{color}
\usepackage{slashed}
\usepackage{xspace}
\usepackage{url}
\usepackage{undertilde}
\usepackage{accents}
\usepackage[makeroom]{cancel}
\usepackage{hyperref}
\usepackage{afterpage}
\usepackage{cite}

\newcommand{\bra}[1]{\langle #1 |}
\newcommand{\ket}[1]{|#1\rangle}
\newcommand{\braket}[2]{\langle #1 | #2 \rangle}

\newcommand{\be}{\begin{equation}}
\newcommand{\ee}{\end{equation}}
\newcommand{\bea}{\begin{eqnarray}}
\newcommand{\eea}{\end{eqnarray}}
\newcommand{\bal}{\begin{align}}
\newcommand{\eal}{\end{align}}

\newcommand{\Eq}{Eq.\@\xspace}
\newcommand{\Eqs}{Eqs.\@\xspace}
\newcommand{\eg}{e.g.\@\xspace}
\newcommand{\ie}{i.e.\@\xspace}
\newcommand{\Fig}{Fig.\@\xspace}
\newcommand{\Figs}{Figs.\@\xspace}

\newcommand{\updown}[2]{^{#1}_{\phantom{#1}#2}}

\newcommand{\half}{\tfrac{1}{2}}

\newcommand{\Id}{\mathbbm{1}}

\newcommand{\C}{\mathbbm{C}}

\newcommand{\makeSymbol}[1]{\mathord{\vcenter{\hbox{#1}}}}

\numberwithin{equation}{section}

\newsavebox{\mybox}

\begin{document}

\begin{center}

\Large
\textbf{Coherent 3$j$-symbol representation for the loop quantum gravity intertwiner space}

\vspace{12pt}

\large
E. Alesci$^{1}$, J. Lewandowski$^{2}$ and I. M\"akinen$^{2}$

\vspace{6pt}

\normalsize
$^1$SISSA, Via Bonomea 265, 34126 Trieste, Italy and INFN Sez. Trieste.\\

$^2$Faculty of Physics, University of Warsaw, ul. Pasteura 5, 02-093 Warsaw, Poland 
\vspace{6pt}

\end{center}

\begin{abstract}

We introduce a new technique for dealing with the matrix elements of the Hamiltonian operator in loop quantum gravity, based on the use of intertwiners projected on coherent states of angular momentum. We give explicit expressions for the projections of intertwiners on the spin coherent states in terms of complex numbers describing the unit vectors which label the coherent states. Operators such as the Hamiltonian can then be reformulated as differential operators acting on polynomials of these complex numbers. This makes it possible to describe the action of the Hamiltonian geometrically, in terms of the unit vectors originating from the angular momentum coherent states, and opens up a way towards investigating the semiclassical limit of the dynamics via asymptotic approximation methods.

\end{abstract}

\section{Introduction}

A new Hamiltonian operator for loop quantum gravity \cite{LQG1,LQG2,LQG3} has been introduced in \cite{LQGsf,newsc}, and some simple matrix elements of the operator have been computed using $SU(2)$ recoupling theory techniques. In this work we propose an alternative method for dealing with such computations, based on intertwiners projected on angular momentum coherent states, in order to overcome the difficulties involved in the use of recoupling theory (see for example \cite{rec1, rec2}). \\

The use of $SU(2)$ coherent states was first introduced in loop quantum gravity in the context of spinfoam models (see \cite{SpinFoam} for a review) with the so called Livine-Speziale coherent intertwiners, introduced in \cite{LivineSpeziale}, studied further in \cite{ConradyFreidel, FKL, Bianchi:2010gc, FS}, and used to compute the asymptotics of the EPRL spin foam model in \cite{EPRL}. In particular, in \cite{ConradyFreidel} the relation between coherent intertwiners and the quantum geometry of a tetrahedron was made explicit by establishing an isomorphism between the four-valent coherent intertwiner and the quantization of the reduced phase space of tetrahedra. These results were further extended to the case of $n$-valent intertwiners with the tools of holomorphic quantization in \cite{FKL}. Further work on intertwiners in the holomorphic representation can be found in \cite{FH, Freidel:2013fia}. \\

The main purpose of the present work is to introduce a way of applying the angular momentum coherent states to computations in canonical loop quantum gravity. We achieve this by rewriting the standard spin network states in terms of representation matrices and intertwiners projected onto the basis of angular momentum coherent states. On the technical level, we will be therefore interested in the object
\be\label{1}
\iota_{\xi_1\cdots\xi_N} = \braket{\iota}{j_1\xi_1\otimes\cdots\otimes j_N\xi_N},
\ee
where $\ket\iota \in {\rm Inv}\,({\cal H}_{j_1}\otimes\cdots\otimes{\cal H}_{j_N})$ is an intertwiner in a standard recoupling theory basis, and $\ket{j\xi}$ are the coherent states of angular momentum, parametrized in terms of a complex number $\xi$ encoding the polar angles of a unit vector $\vec n$. The significance of \eqref{1}, which we evaluate explicitly in the case of three- and four-valent intertwiners, is demonstrated by the following three interpretations of the object $\iota_{\xi_1\cdots\xi_N}$:
\begin{itemize}
\item By definition, $\iota_{\xi_1\cdots\xi_N}$ gives (the complex conjugate of) the projection of the intertwiner $\ket\iota$ onto the coherent state $\ket{j_1\xi_1\otimes\cdots\otimes j_N\xi_N}$ of the space ${\cal H}_{j_1}\otimes\cdots\otimes{\cal H}_{j_N}$. 
\item The components of the Livine--Speziale coherent intertwiner labeled by the vectors $\vec n_{\xi_1},\dots,\vec n_{\xi_N}$ with respect to a standard orthonormal basis $\{\ket{\iota^{(k)}}\}$ of the intertwiner space ${\rm Inv}\,({\cal H}_{j_1}\otimes\cdots\otimes{\cal H}_{j_N})$ are given by the numbers $\iota^{(k)}_{\xi_1\cdots\xi_N}$.
\item Finally, from the explicit expression of $\iota_{\xi_1\cdots\xi_N}$ one can extract a representation of the intertwiner $\ket\iota$ as a polynomial\footnote{This is based on the fact that the spin-$j$ representation of $SU(2)$ can be realized on the space of polynomials of degree $2j$ of a single complex variable.} of the complex variables $\xi_1,\dots,\xi_N$. Furthermore, the standard operators of loop quantum gravity, in particular the Hamiltonian, can be reformulated as differential operators acting on this polynomial.
\end{itemize}
The main motivation for our work is that by reformulating our calculations in the language of the complex numbers $\xi_a$ (as opposed to the language of $SU(2)$ recoupling theory), we will be able express the action of the Hamiltonian on spin network states in a more transparent, geometrical form in terms of the unit vectors $\vec n_a$ corresponding to the complex numbers $\xi_a$. This will hopefully open up a way to investigate the geometrical content and the semiclassical properties of the Hamiltonian \eg by use of saddle point techniques. However, the purpose of the present paper is merely to lay down the framework for carrying out such computations. In particular, obtaining a systematic understanding of the geometrical structure of the Hamiltonian in this framework is a question left for future work. \\

This paper is organized as follows. In section \ref{AMCS} we review the standard construction of intertwiners in loop quantum gravity, and the coherent states of angular momentum, and use these coherent states to rewrite the standard spin network states of LQG in terms of intertwiners projected on the angular momentum coherent states (\ie $\iota_{\xi_1\cdots\xi_N}$). In section \ref{IinCbasis}, we derive explicit expressions for $\iota_{\xi_1\cdots\xi_N}$ in the case of three- and four-valent intertwiners, both in terms of the complex numbers $\xi_a$ and the corresponding unit vectors $\vec n_a$. In section \ref{intpoly} we recall the representation of $SU(2)$ in a space of complex polynomials, and show how the $N$-valent intertwiner $\ket\iota$ can represented as a polynomial of $N$ complex variables $\xi_1,\dots,\xi_N$. In section \ref{LQGapplication}, we give an example of applying the techniques developed in sections \ref{AMCS}--\ref{intpoly} in a concrete calculation. We then close the paper by reporting our conclusions and plans for future work.

\section{Intertwiners and angular momentum coherent states}\label{AMCS}

\subsection{Spin networks and intertwiners in loop quantum gravity}

In loop quantum gravity \cite{LQG1, LQG2, LQG3}, an important role is played by the rotationally invariant subspace of the tensor product ${\cal H}_{j_1}\otimes\cdots\otimes{\cal H}_{j_N}$. There is a one-to-one correspondence between elements of this space, which is often denoted by ${\rm Inv}\,({\cal H}_{j_1}\otimes\cdots\otimes{\cal H}_{j_N})$, and intertwiners, \ie invariant tensors of ${\cal H}_{j_1}\otimes\cdots\otimes{\cal H}_{j_N}$. The Hilbert space of gauge-invariant states in loop quantum gravity is spanned by the so-called spin network states. A spin network state is labeled by a graph $\Gamma$ (oriented, embedded in the three-dimensional spatial manifold $\Sigma$), a spin quantum number $j_e$ for each edge $e$ of $\Gamma$, and an intertwiner $\iota_n \in {\rm Inv}\,\bigl(\bigotimes_{e\in n} {\cal H}_{j_e}\bigr)$ for each node of $\Gamma$, where $\{e\}_{e\in n}$ denotes the set of edges incident on the node $n$. Explicitly, the wave function of a spin network state is given by
\be\label{SN}
\Psi_{\Gamma,\{j_e\},\{\iota_n\}}\bigl(h_{e_1}[A],\dots,h_{e_{N_e}}[A]\bigr) = \biggl(\prod_{e\in\Gamma} \sqrt{d_{j_e}}D^{(j_e)}_{m_en_e}\bigl(h_e[A]\bigr)\biggr) \biggl(\prod_{n\in\Gamma} \iota_n \biggr)_{m_1\cdots m_{N_e};n_1\cdots n_{N_e}}
\ee
Here $N_e$ is the number of edges of the graph $\Gamma$, $h_e[A]\in SU(2)$ is the holonomy of the Ashtekar connection along the edge $e$, and $D^{(j)}_{mn}(h)$ is the representation matrix of $h$ in the spin-$j$ representation of $SU(2)$. The dimension of the representation is denoted by $d_j=2j+1$. The pattern of contraction of the magnetic indices is dictated by the structure of the graph. \\

A basic example of an intertwiner space is ${\rm Inv}\,({\cal H}_{j_1}\otimes{\cal H}_{j_2}\otimes{\cal H}_{j_3})$, the space of three-valent intertwiners. This space is one-dimensional, provided that the spins $j_1$, $j_2$ and $j_3$ satisfy the triangular inequality, and is spanned by the state $\ket{\Psi_0}$, the unique rotationally invariant state of the three spins. The components of this invariant state with respect to the standard basis of ${\cal H}_{j_1}\otimes{\cal H}_{j_2}\otimes{\cal H}_{j_3}$ are given by the Wigner 3$j$-symbol\footnote{In angular momentum theory, we follow the conventions of \cite{Varshalovich}. In particular, we use the Condon--Shortley phase convention for the Clebsch--Gordan coefficients.}:
\be
\ket{\Psi_0} = \sum_{m_1m_2m_3} \begin{pmatrix} j_1&j_2&j_3 \\ m_1&m_2&m_3\end{pmatrix}\,\ket{j_1m_1}\ket{j_2m_2}\ket{j_3m_3}.
\ee
The components of the three-valent intertwiner are therefore given by
\be
\iota^{(j_1j_2j_3)}_{m_1m_2m_3} = \begin{pmatrix} j_1&j_2&j_3 \\ m_1&m_2&m_3\end{pmatrix},
\ee
and invariance of $\iota^{(j_1j_2j_3)}_{m_1m_2m_3}$ under $SU(2)$ follows from the rotational invariance of the state $\ket{\Psi_0}$. Intertwiners of higher valence can be constructed by contracting three-valent intertwiners in a rotationally invariant way, using the tensors $\epsilon^{(j)}_{mn}$ and $\epsilon^{(j)mn}$, both of whose components are defined to be equal to $(-1)^{j-m}\delta_{m,-n}$ with respect to the basis $\{\ket{jm}\}$. For example, the objects
\begin{align}
\iota^{(k)}_{m_1m_2m_3m_4} &= \iota^{(j_1j_2k)}_{m_1m_2\mu}\epsilon^{(k)\mu\nu}\iota^{(kj_3j_4)}_{\nu m_3m_4}\notag \\
&= \sqrt{d_k} \sum_n \begin{pmatrix} j_1&j_2&k \\ m_1&m_2&n\end{pmatrix}(-1)^{k-n}\begin{pmatrix} k&j_3&j_4 \\ -n&m_3&m_4\end{pmatrix} \label{int4}
\end{align}
provide an orthonormal basis of the four-valent intertwiner space ${\rm Inv}\,({\cal H}_{j_1}\otimes{\cal H}_{j_2}\otimes{\cal H}_{j_3}\otimes{\cal H}_{j_4})$, when the intermediate spin $k$ goes through all values compatible with the triangular inequalities.

\subsection{Angular momentum coherent states}\label{spincs}

In this section we will recall some properties of the angular momentum coherent states \cite{Radcliffe, Perelomov}, which will be a central tool in our work. Coherent states of an angular momentum $j$ can be obtained by starting with the state of lowest weight, $\ket{j,-{}j}$, and applying a $SU(2)$ rotation which rotates the vector $\hat e_z = (0,0,1)$ into the vector $\vec n = (\sin\theta\,\cos\phi,\sin\theta\,\sin\phi,\cos\theta)$. The resulting state is denoted by
\be\label{jn}
\ket{j\vec n} = D^{(j)}(\vec n)\ket{j,-{}j}.
\ee
The requirement of rotating the vector $\hat e_z$ into $\vec n$ does not uniquely determine the group element $g(\vec n)$. In order to fix the ambiguity and obtain a unique $g(\vec n)$, one typically specifies that the rotation is performed around an axis in the $xy$-plane. With this choice, the group element effecting the rotation is unique, and is given by
\be\label{gn}
g(\vec n) = e^{-i\theta\vec m\cdot\vec J} = \begin{pmatrix} \cos\theta/2 & -e^{-i\phi}\sin\theta/2 \\ e^{i\phi}\sin\theta/2 & \cos\theta/2 \end{pmatrix},
\ee
where $\vec m = (-\sin\phi,\cos\phi,0)$ is a vector orthogonal to $\vec n$ and lying in the plane spanned by $\hat e_x$ and $\hat e_y$. \\

The states \eqref{jn} form a basis of the Hilbert space ${\cal H}_j$, as indicated by the resolution of identity
\be\label{cs-1}
\Id = d_j\int \frac{d^2n}{4\pi}\,\ket{j\vec n}\bra{j\vec n},
\ee
where $d^2n = \sin\theta\,d\theta\,d\phi$ is the standard measure on the unit sphere. The state $\ket{j\vec n}$ is an eigenstate of the operator $\vec n\cdot\vec J$ with eigenvalue $-j$, just as $\ket{j,-{}j}$ is an eigenstate of $J_z$ with eigenvalue $-j$. \\

For large $j$, the state $\ket{j\vec n}$ has a semiclassical interpretation, describing an angular momentum oriented in the direction $-\vec n$, in the sense that the expectation value of the angular momentum in the state is given by $\bra{j\vec n}\vec J\ket{j\vec n} = -j\vec n$, and the relative uncertainty $\Delta J/\langle J\rangle = {\cal O}(j^{-1/2})$ is small in the limit of large $j$. Another useful property of the states \eqref{jn} is that the spin-$j$ coherent state can be constructed as a simple tensor product of spin-1/2 coherent states:
\be\label{jn-product}
\ket{j\vec n} = \underbrace{\ket{\vec n}\otimes\cdots\otimes\ket{\vec n}}_{\text{2$j$ times}}
\ee
where $\ket{\vec n}\equiv \ket{\half,\vec n}$ denotes the coherent state in the spin-1/2 representation. \\

For practical purposes, it is convenient to introduce the complex parameter
\be
\xi = -e^{-i\phi}\tan\frac{\theta}{2},
\ee
which encodes the angles of the vector $\vec n$ and can therefore be used instead of $\vec n$ to label the coherent states \eqref{jn}. Using the decomposition of a $SU(2)$ group element given by \Eq \eqref{+0-dec} of Appendix \ref{app:su2}, we find that we can write the group element \eqref{gn} in terms of $\xi$ as
\be\label{g+0-}
g(\xi) = e^{\xi J_+}e^{\ln(1+|\xi|^2)J_0}e^{-\bar\xi J_-}.
\ee
The explicit matrix form of this group element reads
\be\label{gxi}
g(\xi) = \frac{1}{\sqrt{1+|\xi|^2}}\begin{pmatrix} 1&\xi \\ -\bar\xi&1 \end{pmatrix}.
\ee
By applying \eqref{g+0-} to the state $\ket{j,-{}j}$, we obtain the expression
\be\label{jxi}
\ket{j\xi} = \frac{1}{(1+|\xi|^2)^j}e^{\xi J_+}\ket{j,-{}j}
\ee
for the coherent state \eqref{jn}. Furthermore, in terms of $\xi$ the resolution of identity \eqref{cs-1} takes the form
\be\label{xi-1}
\Id = \int d\mu_j(\xi)\,\ket{j\xi}\bra{j\xi},
\ee
the measure being given by
\be\label{dmu_j}
d\mu_j(\xi) = \frac{d_j}{\pi}\frac{d^2\xi}{(1+|\xi|^2)^2}.
\ee

\Eqs \eqref{jxi} and \eqref{gxi} are very useful in practical calculations. For example, using \eqref{gxi} together with \eqref{jn-product}, the scalar product between two coherent states is immediately calculated as
\be\label{xieta}
\braket{j\xi}{j\eta} = \bra - g^\dagger(\xi)g(\eta)\ket -^{2j} = \frac{(1+\bar\xi\eta)^{2j}}{(1+|\xi|^2)^j(1+|\eta|^2)^j},
\ee
with $\ket - \equiv \ket{\half,{-}\half}$ the state of lowest weight in the spin-1/2 representation. In terms of the vectors $\vec n$, one has
\be\label{mn}
\braket{j\vec m}{j\vec n} = e^{ij\alpha(\vec m,\vec n)}\biggl(\frac{1+\vec m\cdot\vec n}{2}\biggr)^j,
\ee
where
\be\label{area}
\alpha(\vec m,\vec n) = \ln\frac{1+\overline{\xi_{\vec m}}\xi_{\vec n}}{1+\xi_{\vec m}\overline{\xi_{\vec n}}}
\ee
is the (oriented) area of the spherical triangle spanned by the vectors $\hat e_z$, $\vec m$ and $\vec n$. \\

To give another example, let us compute the matrix elements of the angular momentum operator between the states $\ket{j\xi}$. In the calculation, it is more convenient to use the unnormalized states
\be\label{jxi-unnorm}
|j\xi) = e^{\xi J_+}\ket{j,-{}j},
\ee
for which we have
\be
(j\xi|j\eta) = (1+\bar\xi\eta)^{2j},
\ee
and from which the coherent states $\ket{j\xi}$ are obtained by multiplying with the appropriate normalization factor,
\be
\ket{j\xi} = \frac{1}{(1+|\xi|^2)^j}|j\xi).
\ee
To find, for example, the matrix element of $J_+$, we consider
\be
(j\xi|e^{\epsilon J_+}|j\eta) = (j\xi|e^{\epsilon J_+}e^{\eta J_+}\ket{j,-{}j} = (j\xi|j,\eta+\epsilon).
\ee
Expanding to linear order in $\epsilon$, we get
\be
(j\xi|J_+|j\eta) = \frac{\partial}{\partial\eta}(j\xi|j\eta) = 2j\bar\xi(1+\bar\xi\eta)^{2j-1}.
\ee
A similar calculation for $J_-$ shows that
\be
(j\xi|J_-|j\eta) = \frac{\partial}{\partial\bar\xi}(j\xi|j\eta) = 2j\eta(1+\bar\xi\eta)^{2j-1}.
\ee
Finding the matrix element of $J_0$ is less straightforward, but can be done by considering the identity $e^{\eta J_+}J_-e^{-\eta J_+}|j\eta) = 0$ \cite{Perelomov}. Using the Baker-Campbell-Hausdorff formula, one obtains $(J_- + 2\eta J_0 - \eta^2 J_+)|j\eta)=0$, from which it follows that
\be
(\xi|J_0|\eta) = j(\bar\xi\eta-1)(1+\bar\xi\eta)^{2j-1}.
\ee
Summarizing our findings, and rewriting them in terms of the normalized states $\ket{j\xi}$, we have
\begin{align}
\bra{j\xi}J_+\ket{j\eta} &= 2j\bar\xi\frac{(1+\bar\xi\eta)^{2j-1}}{(1+|\xi|^2)^j(1+|\eta|^2)^j} \label{J+xy}\\
\bra{j\xi}J_0\ket{j\eta} &= j(\bar\xi\eta-1)\frac{(1+\bar\xi\eta)^{2j-1}}{(1+|\xi|^2)^j(1+|\eta|^2)^j} \label{J0xy} \\
\bra{j\xi}J_-\ket{j\eta} &= 2j\eta\frac{(1+\bar\xi\eta)^{2j-1}}{(1+|\xi|^2)^j(1+|\eta|^2)^j}. \label{J-xy}
\end{align}

To conclude our discussion of the angular momentum coherent states, let us note some useful properties of the matrix $g(\xi)$. Firstly, the inverse matrix is obtained by inverting the sign of the parameter, \ie
\be\label{g-1}
g^{-1}(\xi) = g(-\xi).
\ee
Secondly, if $\xi$ is the parameter associated with the vector $\vec n$, then the parameter associated with the vector $-\vec n$ is $-1/\bar\xi$; thus
\be
g(-\vec n) = g\bigl(-\bar\xi_{\vec n}^{-1}\bigr).
\ee
Furthermore, by using the explicit matrix representation \eqref{gxi}, one can derive the composition law of the group elements $g(\xi)$. The result reads
\be\label{g1g2}
g(\xi_1)g(\xi_2) = g(\xi_{12})e^{i\alpha_{12}\sigma_z/2},
\ee
where
\be
\xi_{12} = \frac{\xi_1+\xi_2}{1-\bar\xi_1\xi_2} \qquad \text{and} \qquad e^{i\alpha_{12}/2} = \frac{1-\xi_1\bar\xi_2}{|1-\xi_1\bar\xi_2|}.
\ee
Note that here $\alpha_{12}$ is again the area of a spherical triangle, as given by \Eq \eqref{area}.

\subsection{Livine--Speziale coherent intertwiners}\label{LS-sect}

The spin coherent states described in the previous section can be used to construct an alternative basis of the intertwiner space ${\rm Inv}\,({\cal H}_{j_1}\otimes\cdots\otimes{\cal H}_{j_N})$. This basis is formed by the coherent intertwiners introduced by Livine and Speziale \cite{LivineSpeziale}. A coherent intertwiner is constructed by forming a tensor product of the spin coherent states \eqref{jn}, and making the resulting state rotationally invariant by applying a group averaging:
\be
\ket{j_1\cdots j_N;\vec n_1\cdots\vec n_N} = \int dg\,D^{(j_1)}(g)\ket{j_1\vec n_1}\otimes \cdots \otimes D^{(j_N)}(g)\ket{j_N\vec n_N}.
\ee
The coherent intertwiners $\ket{j_1\cdots j_N;\vec n_1\cdots\vec n_N}$ provide an overcomplete basis in the space of $N$-valent intertwiners. Physically, the Livine--Speziale intertwiners were first recognized to represent semiclassical polyhedra in the same sense in which an ordinary intertwiner describes a quantum polyhedron \cite{vol1, vol2, Barbieri, Bianchi:2010gc}, and the description of the kinematical states of LQG in terms of coherent intertwiners led to the concept of twisted geometries \cite{FS} (\ie discrete geometries different from the usual Regge geometries, obtained by matching the normal vectors corresponding to faces of classical polyhedra, with shapes of the glued faces not necessarily matching.) \\

For the purposes of the present work, we are interested in how an $N$-valent coherent intertwiner is expanded in some standard basis of ${\rm Inv}\,({\cal H}_{j_1}\otimes\cdots\otimes{\cal H}_{j_N})$, for example one constructed by extending \Eq \eqref{int4} to the $N$-valent case. To this end, consider
\be\label{LS-component}
\braket{j_1m_1\otimes \cdots \otimes j_Nm_N}{j_1\cdots j_N;\vec n_1\cdots\vec n_N} = \int dg\,\prod_{a=1}^N \bra{j_am_a}D^{(j_a)}(g)D^{(j_a)}(\vec n_a)\ket{j_a,-{}j_a}.
\ee
On the right we can insert identity in the form $\Id = \sum_{m_a} \ket{j_am_a}\bra{j_am_a}$ between the two $D$-matrices. The resulting integral over the group is then given by\footnote{To see that this equation is valid, note that both sides of the equation are a representation of the projection operator onto ${\rm Inv}\,({\cal H}_{j_1}\otimes\cdots\otimes{\cal H}_{j_N})$.}
\be
\int dg\,D^{(j_1)}_{m_1n_1}(g)\cdots D^{(j_N)}_{m_Nn_N}(g) = \sum_\iota \iota_{m_1\cdots m_N}\iota_{n_1\cdots n_N},
\ee
where the sum on the right runs through any real\footnote{In case the basis is not real, the intertwiner $\iota_{m_1\cdots m_N}$ on the right-hand side should be complex conjugated.}, orthonormal basis of the intertwiner space ${\rm Inv}\,({\cal H}_{j_1}\otimes\cdots\otimes{\cal H}_{j_N})$. From \Eq \eqref{LS-component}, we therefore obtain
\begin{align}
&\braket{j_1m_1\otimes \cdots \otimes j_Nm_N}{j_1\cdots j_N;\vec n_1\cdots\vec n_N} \notag \\
&= \sum_\iota\Bigr(\sum_{n_1\cdots n_N} D^{(j_1)}_{n_1,-{}j_1}(\vec n_1)\cdots D^{(j_N)}_{n_N,-{}j_N}(\vec n_N)\iota_{n_1\cdots n_N}\Bigr)\iota_{m_1\cdots m_N}.
\end{align}
From this we see that the decomposition of the coherent intertwiner in the basis $\{\ket\iota\}$ reads
\be
\ket{j_1\cdots j_N;\vec n_1\cdots\vec n_N} = \sum_\iota c_{\vec n_1\cdots\vec n_N}(\iota) \ket\iota,
\ee
where the projections of $\ket{j_1\cdots j_N;\vec n_1\cdots\vec n_N}$ onto the basis intertwiners are given by
\be\label{c(i)}
c_{\vec n_1\cdots\vec n_N}(\iota) = \sum_{n_1\cdots n_N} \iota_{n_1\cdots n_N} D^{(j_1)}_{n_1,-{}j_1}(\vec n_1)\cdots D^{(j_N)}_{n_N,-{}j_N}(\vec n_N).
\ee

\subsection{Spin networks in the coherent-state basis}

The starting point for our work in the remainder of this paper is to express the spin network state of \Eq \eqref{SN} using the basis of angular momentum coherent states, as opposed to the standard basis of magnetic indices, which is used in \Eq \eqref{SN}. If one starts with the expression \eqref{SN}, but uses in each Hilbert space ${\cal H}_j$ the resolution of identity $\Id = \int d\mu_j(\xi)\,\ket{j\xi}\bra{j\xi}$ instead of $\Id = \sum_m \ket{jm}\bra{jm}$, one obtains the following expression for the spin network $\Psi_{\Gamma,\{j_e\},\{\iota_n\}}$: 
\begin{align}
\Psi_{\Gamma,\{j_e\},\{\iota_n\}}\bigl(h_{e_1}[A],\dots,h_{e_{N_e}}[A]\bigr) &= \int d\mu_{j_1}(\xi_1)\,d\mu_{j_1}(\eta_1)\cdots d\mu_{j_{N_e}}(\xi_{N_e})\,d\mu_{j_{N_e}}(\eta_{N_e}) \notag \\
&{}\times\biggl(\prod_{e\in\Gamma} \sqrt{d_{j_e}}D^{(j_e)}_{\xi_e\eta_e}\bigl(h_e[A]\bigr)\biggr) \biggl(\prod_{n\in\Gamma} \iota_n \biggr)_{\xi_1\cdots\xi_{N_e};\eta_1\cdots\eta_{N_e}}\label{csSN}.
\end{align}
Here $D^{(j)}_{\xi\eta}(h)$ denotes the matrix elements of the Wigner matrix $D^{(j)}(h)$ with respect to the basis of coherent states, \ie
\be
D^{(j)}_{\xi\eta}(h) = \bra{j\xi}D^{(j)}(h)\ket{j\eta}.
\ee
Similarly, the intertwiners in \Eq \eqref{csSN} are constructed by projecting each state $\ket{\iota_n}$ onto the coherent-state basis of the space $\bigotimes_{e\in n} {\cal H}_{j_{e_n}}$. A detailed discussion of the objects so obtained, $\braket{j_1\xi_1\otimes\cdots\otimes j_N\xi_N}{\iota}$, is the subject of the following section.

\section{Intertwiners in the basis of spin coherent states}
\label{IinCbasis}

We will now consider in more detail the object
\begin{align}
\iota_{\xi_1\cdots\xi_N} &\equiv \braket{\iota}{j_1\xi_1\otimes\cdots\otimes j_N\xi_N} \notag \\
&= \iota_{m_1\cdots m_N} D^{(j_1)}_{m_1,-{}j_1}(\xi_1)\cdots D^{(j_N)}_{m_N,-{}j_N}(\xi_N), \label{iotaxi}
\end{align}
which appears implicitly in \Eq \eqref{csSN}, and which by definition is the (complex conjugate of) the component of the intertwiner $\ket\iota$ with respect to the coherent state basis $\ket{j_1\xi_1}\cdots\ket{j_N\xi_N}$ of ${\cal H}_{j_1}\otimes\cdots\otimes{\cal H}_{j_N}$. Moreover, as shown by our discussion in section \ref{LS-sect}, $\iota_{\xi_1\cdots\xi_N}$ is also the component of the Livine--Speziale intertwiner $\ket{j_1\cdots j_N;\xi_1\cdots\xi_N}$ with respect to the basis element $\ket\iota$ of the intertwiner space ${\rm Inv}\,({\cal H}_{j_1}\otimes\cdots\otimes{\cal H}_{j_N})$. A third interpretation of our result for $\iota_{\xi_1\cdots\xi_N}$ will be discussed in section \ref{intpoly}. \\

Below we will perform the evaluation of \eqref{iotaxi} explicitly in the case of three- and four-valent intertwiners.

\subsection{The three-valent intertwiner}

We start with the calculation of
\be\label{iDDD}
\iota_{\xi_1\xi_2\xi_3} \equiv \begin{pmatrix} j_1&j_2&j_3 \\ \xi_1&\xi_2&\xi_3 \end{pmatrix} = \begin{pmatrix} j_1&j_2&j_3 \\ m_1&m_2&m_3 \end{pmatrix}D^{(j_1)}_{m_1,-{}j_1}(\xi_1)D^{(j_2)}_{m_2,-{}j_2}(\xi_2)D^{(j_3)}_{m_3,-{}j_3}(\xi_3),
\ee
which is conveniently carried out by using the realization of the spin-$j$ representation of $SU(2)$ as a symmetrized tensor product of $2j$ copies of the fundamental representation, as described in Appendix \ref{su2prod}. A magnetic index $m$ in the spin-$j$ representation corresponds to a symmetrized set of $2j$ spinor indices, $(A_1\cdots A_{2j})$. The components of the three-valent intertwiner in the magnetic basis are then given by
\be\label{iABC}
\iota_{(A_1\cdots A_{2j_1})(B_1\cdots B_{2j_2})(C_1\cdots C_{2j_3})} = N_{j_1j_2j_3}I_{(A_1\cdots A_{2j_1})(B_1\cdots B_{2j_2})(C_1\cdots C_{2j_3})},
\ee
where
\be
I_{A_1\cdots A_{2j_1}B_1\cdots B_{2j_2}C_1\cdots C_{2j_3}} = \epsilon_{A_1B_1}\cdots\epsilon_{A_aB_a}\epsilon_{B_{a+1}C_1}\cdots\epsilon_{B_{a+b}C_b}\epsilon_{C_{b+1}A_{a+1}}\cdots\epsilon_{C_{b+c}A_{a+c}},
\ee
and $a = j_1+j_2-j_3, b = j_2+j_3-j_1, c = j_3+j_1-j_2$, and the normalization factor $N_{j_1j_2j_3}$ is given by \Eq \eqref{Njjj}. The representation matrix $D^{(j)}_{mn}(g)$ has the form \eqref{Djprod}; in particular, when the second index takes its minimal value, as in \Eq \eqref{iDDD}, we have 
\be\label{DA-}
D^{(j)}_{m,-{}j}(g) = D^{(j)}_{(A_1\cdots A_{2j})(-...-)}(g) = g^{A_1}_{\quad -} \cdots g^{A_{2j}}_{\quad -}.
\ee
Using \Eqs \eqref{iABC} and \eqref{DA-} in \Eq \eqref{xi3j}, we get
\begin{align}
\begin{pmatrix} j_1&j_2&j_3 \\ \xi_1&\xi_2&\xi_3 \end{pmatrix} = N_{j_1j_2j_3}(\epsilon_{\xi_1\xi_2})^{j_1+j_2-j_3}(\epsilon_{\xi_2\xi_3})^{j_2+j_3-j_1}(\epsilon_{\xi_3\xi_1})^{j_3+j_1-j_2},
\end{align}
where we introduced the notation
\be
\epsilon_{\xi\eta} = \epsilon_{AB}g^A_{\quad-}(\xi)g^B_{\quad-}(\eta).
\ee
Recalling \Eq \eqref{gxi}, and choosing $\epsilon_{+-} = +1$, we get
\be
\epsilon_{\xi\eta} = \frac{\xi-\eta}{\sqrt{1+|\xi|^2}\sqrt{1+|\eta|^2}}.
\ee
Therefore we find that the coherent components of the 3$j$-symbol are given by
\be\label{xi3j}
\begin{pmatrix} j_1&j_2&j_3 \\ \xi_1&\xi_2&\xi_3 \end{pmatrix} = N_{j_1j_2j_3}\frac{(\xi_1-\xi_2)^{j_1+j_2-j_3}(\xi_2-\xi_3)^{j_2+j_3-j_1}(\xi_3-\xi_1)^{j_3+j_1-j_2}}{(1+|\xi_1|^2)^{j_1}(1+|\xi_2|^2)^{j_2}(1+|\xi_3|^2)^{j_3}}.
\ee
An alternative derivation of this result, using the complex polynomial representation of $SU(2)$ (see Appendix \ref{sec:Pj}) is reported in Appendix \ref{app:poly}. 

\subsection{Four-valent intertwiners}

In order to calculate the components of the four-valent intertwiner in the coherent state basis, it is convenient to use the graphical representation given by \Eq \eqref{i4graphic}, and reproduced below:
\be\label{i4fig}
\iota^{(k)}_{(A_1\cdots A_{2j_1})\cdots(D_1\cdots D_{2j_4})} = \sqrt{d_k}N_{j_1j_2k}N_{kj_3j_4}\;\makeSymbol{\includegraphics[scale=0.75]{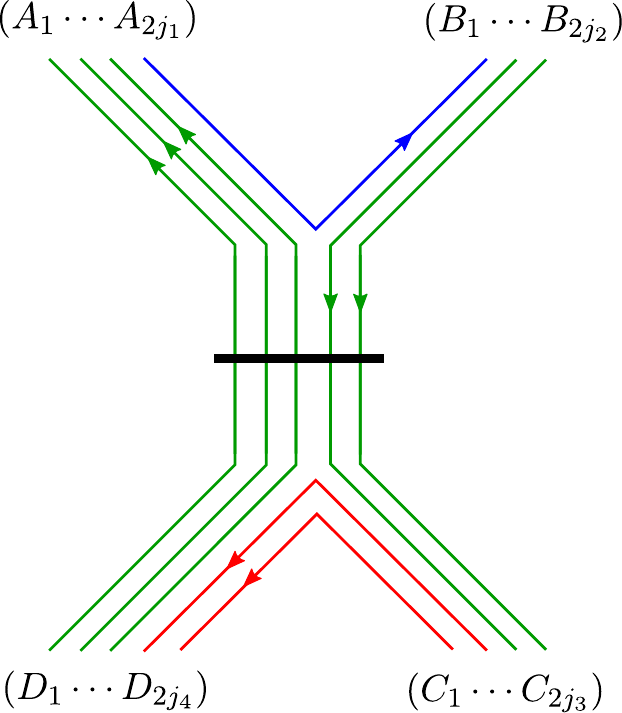}}
\ee
When $\iota^{(k)}_{m_1\cdots m_4}$ is contracted with $D^{(j_1)}_{m_1,-{}j_1}(\xi_1)\cdots D^{(j_4)}_{m_4,-{}j_4}(\xi_4)$ in order to produce $\iota^{(k)}_{\xi_1\cdots\xi_4}$, each blue line in \Eq \eqref {i4fig} becomes $\epsilon_{\xi_1\xi_2}$, and each red line becomes $\epsilon_{\xi_3\xi_4}$. The green lines (which are symmetrized over the internal group of lines) become
\be
{\cal N}\sum_{n_{ab}} N(n_{ab})(\epsilon_{\xi_3\xi_1})^{n_{13}}(\epsilon_{\xi_4\xi_1})^{n_{14}}(\epsilon_{\xi_2\xi_3})^{n_{23}}(\epsilon_{\xi_2\xi_4})^{n_{24}},
\ee
where $n_{ab}$ denotes the number of lines connecting external spins $j_a$ and $j_b$, and $N(n_{ab})$ is the number of ways of arranging the lines for fixed values of the $n_{ab}$. Furthermore, ${\cal N}$ is an overall normalization factor (ensuring that the symmetrization is done with total weight 1), and the sum runs over those values of $n_{ab}$ that satisfy the constraints
\begin{align}
n_{13} + n_{14} &= j_1-j_2+k, \label{N_1}\\
n_{23} + n_{24} &= j_2-j_1+k, \\
n_{13} + n_{23} &= j_3-j_4+k, \\
n_{14} + n_{24} &= j_4-j_3+k. \label{N_4}
\end{align}
These conditions reduce the summation to one independent sum. Determining $N(n_{ab})$ and ${\cal N}$ is a straightforward combinatorial problem, and in the end we obtain the result
\begin{align}
\iota^{(k)}_{\xi_1\cdots\xi_4} = \sqrt{d_k}N_{j_1j_2k}N_{kj_3j_4}\frac{(\xi_1-\xi_2)^{j_1+j_2-k}(\xi_3-\xi_4)^{j_3+j_4-k}Q_k(\xi_1,\xi_2,\xi_3,\xi_4)}{(1+|\xi_1|^2)^{j_1}(1+|\xi_2|^2)^{j_2}(1+|\xi_3|^2)^{j_3}(1+|\xi_4|^2)^{j_4}},\label{xi-int4}
\end{align}
where the function $Q_k$ is given by
\begin{align}
&Q_k(\xi_1,\xi_2,\xi_3,\xi_4) = \dfrac{1}{\begin{pmatrix} 2k \\ k+j_1-j_2 \end{pmatrix}} \sum_t \begin{pmatrix} j_3-j_4+k \\ t \end{pmatrix} \begin{pmatrix} j_4-j_3+k \\ j_1-j_2+k-t \end{pmatrix} \notag \\
&{}\times (\xi_3-\xi_1)^t (\xi_2-\xi_4)^{j_2-j_1-j_3+j_4+t} (\xi_4-\xi_1)^{j_1-j_2+k-t} (\xi_2-\xi_3)^{j_3-j_4+k-t}
\end{align}
(the sum runs over all values of $t$ for which the factorials in the binomial coefficients have non-negative arguments). \\

The method applied above to the four-valent intertwiner can, at least in principle, be extended to obtain the components of an intertwiner of arbitrary valence with respect to the coherent state basis of ${\cal H}_{j_1}\otimes\cdots{\cal H}_{j_N}$. From this it is clear, even without carrying out the calculation in detail, that the $N$-valent intertwiner $\iota_{\xi_1\cdots\xi_N}$ always has the general structure
\be\label{iN_xi}
\iota_{\xi_1\cdots\xi_N} = \frac{P_\iota(\xi_1,\dots,\xi_N)}{(1+|\xi_1|^2)^{j_1}\cdots(1+|\xi_N|^2)^{j_N}},
\ee
where the function $P_\iota(\xi_1,\cdots,\xi_N)$ is a polynomial of order $2j_a$ in each variable $\xi_a$, and is a function of only the differences $\xi_a-\xi_b$. 

\subsection{Expression in terms of normal vectors}\label{normals}

The expression \eqref{iN_xi} for the coherent components of an intertwiner can be rewritten in terms of the vectors $\vec n_a$ corresponding to the parameters $\xi_a$ as follows. We begin by writing $\xi_a-\xi_b$ as $\xi_a(1+\check{\bar\xi}_a\xi_b)$, where $\check\xi_a = -\bar\xi_a^{-1}$ denotes the parameter corresponding to the vector $-\vec n_a$. Next, comparing \Eqs \eqref{xieta} and \eqref{mn} for the scalar product of two coherent states, we see that
\be
1+\bar\xi\eta = e^{i\alpha(\vec m,\vec n)}\sqrt{1+|\xi|^2}\sqrt{1+|\eta|^2}\sqrt{\frac{1+\vec m\cdot\vec n}{2}},
\ee
where $\vec m$ and $\vec n$ are the vectors corresponding to $\xi$ and $\eta$ respectively, and the phase $\alpha(\vec m,\vec n)$ is given by \Eq \eqref{area}. Using this to rewrite the factor $(1+\check{\bar\xi}_a\xi_b)$, we obtain
\be\label{xa-xb}
\xi_a-\xi_b = -e^{i[\alpha(-\vec n_a,\vec n_b)-\phi_a]}\sqrt{1+|\xi_1|^2}\sqrt{1+|\xi_2|^2}\sqrt{\frac{1-\vec n_a\cdot\vec n_b}{2}}.
\ee
As an example, let us consider the 3$j$-symbol given by \Eq \eqref{xi3j}. Making the replacement indicated by \Eq \eqref{xa-xb} in \Eq \eqref{xi3j}, we obtain the following expression for the components of the 3$j$-symbol with respect to the state $\ket{j_1\vec n_1}\ket{j_2\vec n_2}\ket{j_3\vec n_3}$.
\begin{align}
\begin{pmatrix} j_1&j_2&j_3 \\ \vec n_1&\vec n_2&\vec n_3 \end{pmatrix} &= N_{j_1j_2j_3}e^{i(j_1+j_2-j_3)[\alpha(-\vec n_1,\vec n_2)-\phi_1] + i(j_2+j_3-j_1)[\alpha(-\vec n_2,\vec n_3)-\phi_2] + i(j_3+j_1-j_2)[\alpha(-\vec n_3,\vec n_1)-\phi_3]} \notag \\
&{}\times \biggl(\frac{1-\vec n_1\cdot\vec n_2}{2}\biggr)^{\frac{j_1+j_2-j_3}{2}}\biggl(\frac{1-\vec n_2\cdot\vec n_3}{2}\biggr)^{\frac{j_2+j_3-j_1}{2}}\biggl(\frac{1-\vec n_3\cdot\vec n_1}{2}\biggr)^{\frac{j_3+j_1-j_2}{2}}.\label{n3j}
\end{align}
The norm of this object is peaked on configurations where the closure condition $j_1\vec n_1+j_2\vec n_2+j_3\vec n_3=0$ is satisfied, and the peak becomes sharper with increasing values of the spins $j_1$, $j_2$ and $j_3$. This is a special case of the result obtained by Livine and Speziale \cite{LivineSpeziale} on the norm of the coherent intertwiner, since according to the discussion in section \ref{LS-sect}, the norm of \eqref{n3j} has an alternative interpretation as the norm of the three-valent coherent intertwiner $\ket{j_1j_2j_3;\vec n_1\vec n_2\vec n_3}$. We nevertheless illustrate this property numerically in \Figs \ref{3jplot1} and \ref{3jplot2} below. \\

In \Fig \ref{3jplot1}, we plot the norm of \eqref{n3j} as a function of the polar angles of the vector $\vec n_3$, when the first two vectors are fixed as $\vec n_1=(0,0,1)$ and $\vec n_2 = (\tfrac{\sqrt{3}}{2},0,-\tfrac{1}{2})$, and the spins $j_1$, $j_2$ and $j_3$ are equal to a common value $j$. We consider the values $j=20$ and $j=100$, and see that in both cases, the norm is peaked on $\vec n_3=-\vec n_1-\vec n_2$, corresponding to the angles $(\theta,\phi) = (\tfrac{2}{3}\pi,\pi)$, but the peak is sharper in the latter case. In \Fig \ref{3jplot2}, we show the norm of \eqref{n3j} when $j_1=j_2=50$ and $j_3=86\approx 50\sqrt{2}$, and we have fixed $\vec n_1=(1,0,0)$ and $\vec n_2=(0,1,0)$. The norm is peaked on the approximately closed configuration $\vec n_3=\tfrac{1}{\sqrt 2}(-1,-1,0)$, corresponding to the angles $(\theta,\phi) = (\tfrac{1}{2}\pi,\tfrac{5}{4}\pi)$.

\begin{figure}[p]
	\centering
		\includegraphics[width=1\textwidth]{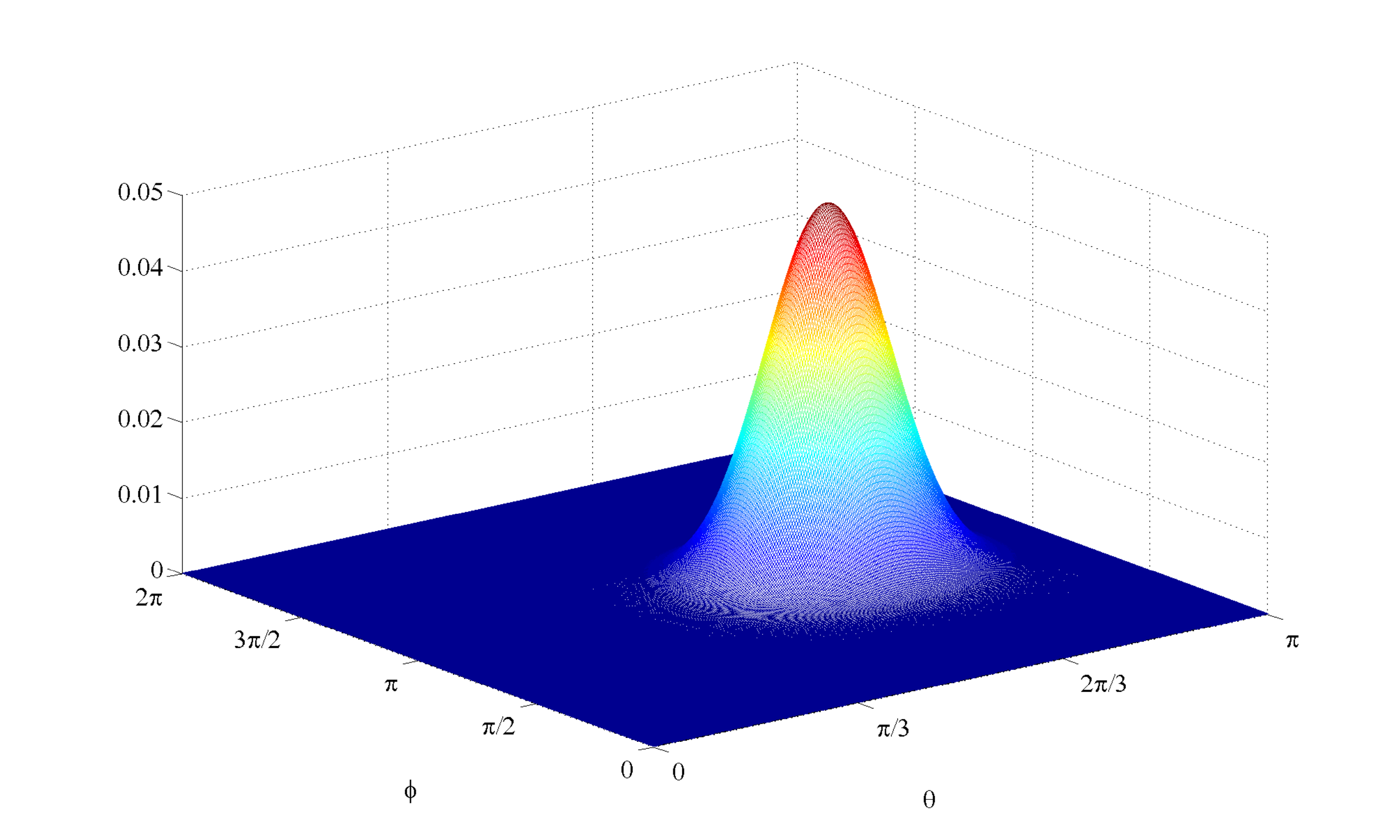}
		\vspace{24pt}
		\includegraphics[width=1\textwidth]{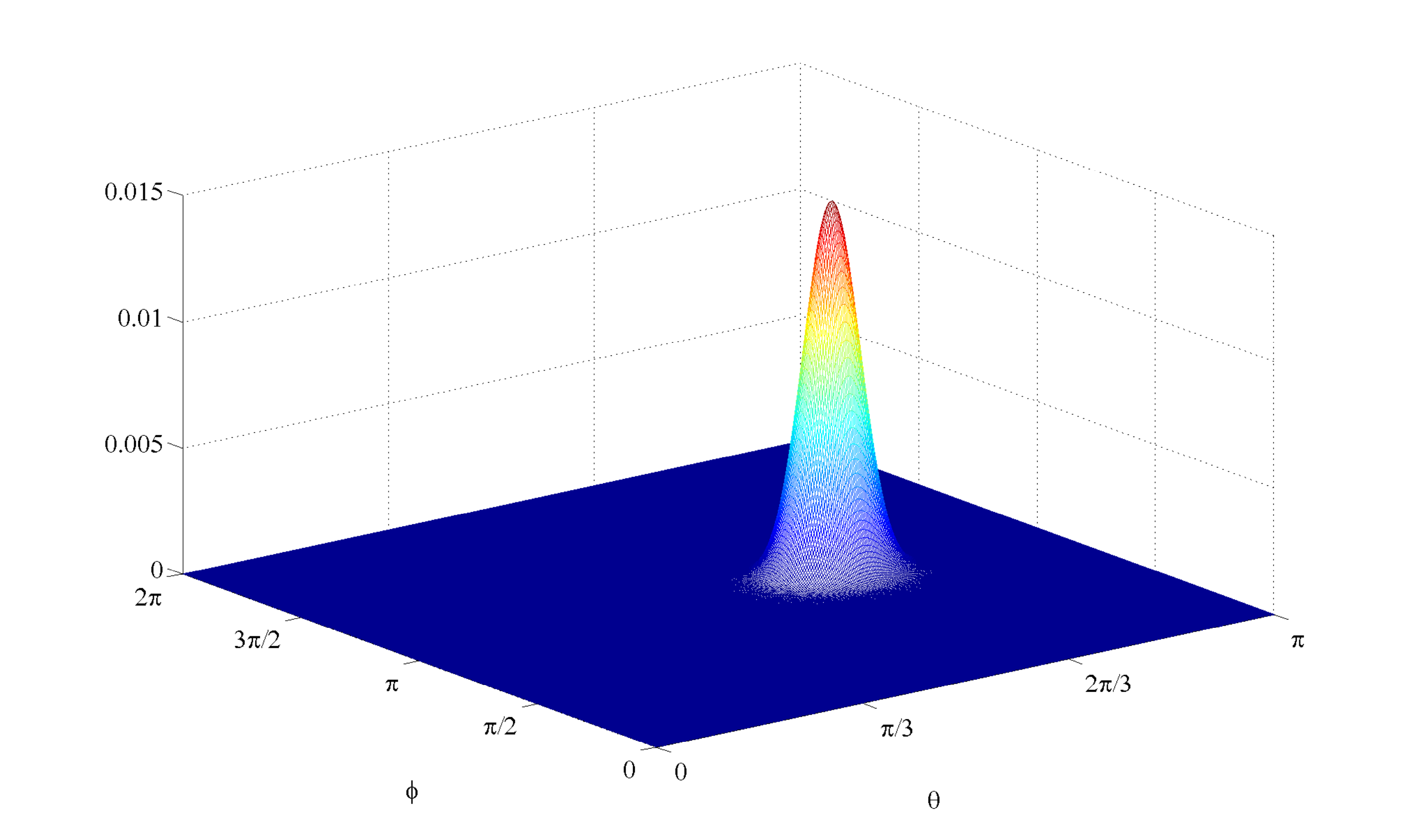}
		\caption{The norm of the 3$j$-symbol \eqref{n3j} as a function of the polar angles of the vector $\vec n_3$. The spins $j_1$, $j_2$ and $j_3$ are all equal to a common value $j$, and the first two vectors are fixed to $\vec n_1=(0,0,1)$ and $\vec n_2 = (\tfrac{\sqrt{3}}{2},0,-\tfrac{1}{2})$. In the upper diagram $j=20$, and in the lower diagram $j=100$. In both cases, the norm is peaked on the angles $(\theta,\phi) = (\tfrac{2}{3}\pi,\pi)$, corresponding to the closed configuration $\vec n_3 = (-\tfrac{\sqrt{3}}{2},0,-\tfrac{1}{2}) = -\vec n_1-\vec n_2$, but the peak is sharper when the spin $j$ is larger.}
	\label{3jplot1}
\end{figure}

\begin{figure}[h]
	\centering
		\includegraphics[width=1\textwidth]{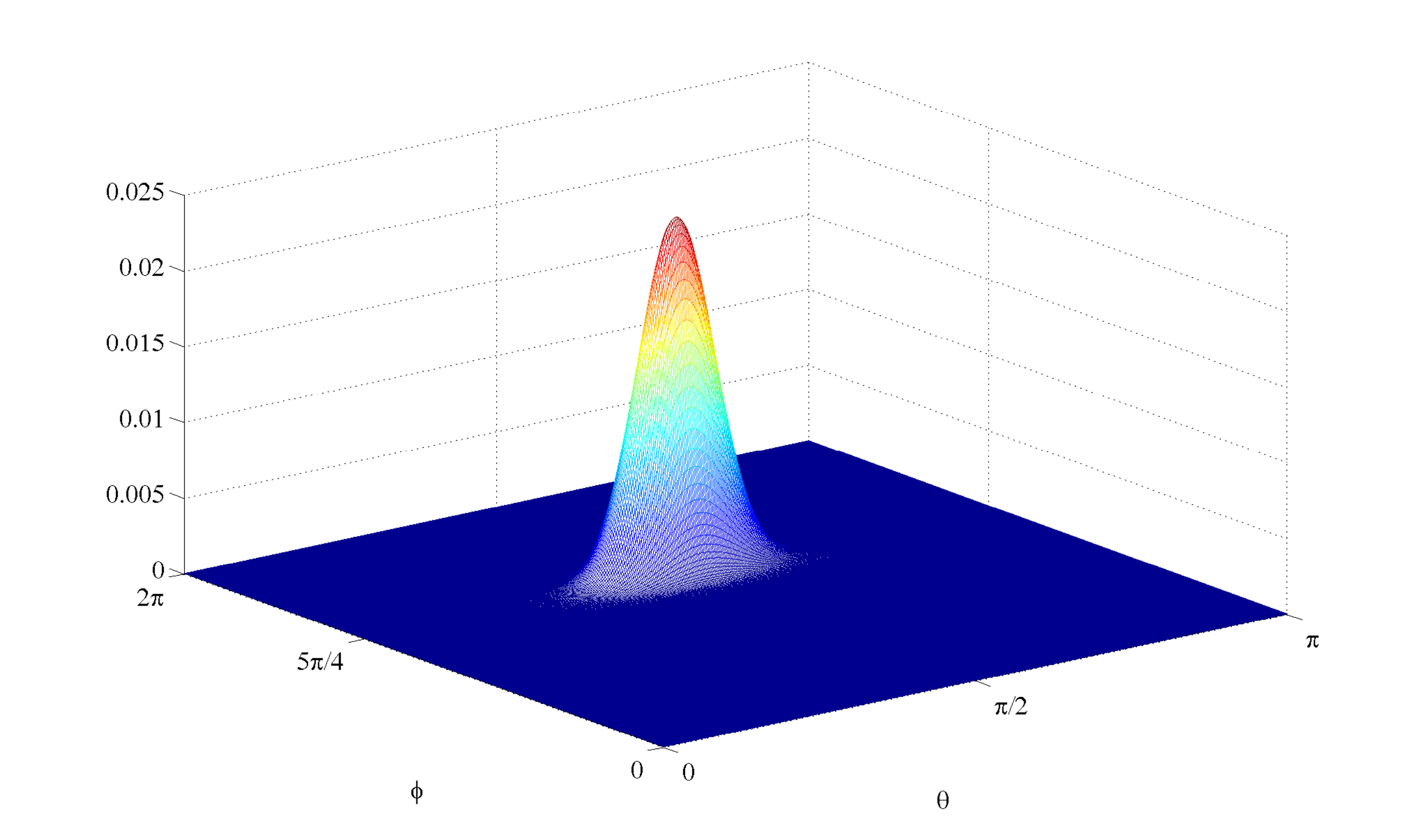}
		\caption{The norm of \eqref{n3j} as a function of the polar angles of $\vec n_3$ when $j_1=j_2=50$ and $j_3=86\approx 50\sqrt{2}$. The maximum occurs at the angles $(\theta,\phi) = (\tfrac{1}{2}\pi,\tfrac{5}{4}\pi)$, corresponding to the vector $\vec n_3 = (-\tfrac{1}{\sqrt 2},{-\frac{1}{\sqrt 2}},0)$. Therefore the norm is peaked on the approximately closed configuration $\vec n_1+\vec n_2+\sqrt{2}\vec n_3=0$.}
	\label{3jplot2}
\end{figure}

\subsection{Raising and lowering indices in the coherent-state basis}

Before concluding this section, we must discuss how indices of intertwiners are raised and lowered in the basis of coherent states. This is necessary for a consistent contraction of indices in an equation such as \eqref{csSN}. In the basis of magnetic indices, indices are raised and lowered using the epsilon tensor; for example,
\be\label{raising-m}
\iota\,\updown{m_1}{m_2\cdots m_N} = \epsilon^{(j_1)m_1n_1}\iota_{n_1m_2\cdots m_N}.
\ee
To derive the rule for raising an index in the coherent-state basis, we can compute the component of
\be
\sum_{m_1\cdots m_N} \iota\,\updown{m_1}{m_2\cdots m_N}\,\bra{j_1m_1}\otimes\ket{j_2m_2}\otimes\cdots\otimes\ket{j_Nm_N}
\ee
with respect to the state $\bra{j_1\xi_1}\otimes\ket{j_2\xi_2}\otimes\cdots\otimes\ket{j_N\xi_N}$, or more directly consider the integral
\be\label{raising-xi}
\iota\,\updown{\xi_1}{\xi_2\cdots\xi_N} = \int d\mu_{j_1}(\eta_1)\,\epsilon^{(j_1)\xi_1\eta_1}\iota_{\eta_1\xi_2\cdots\xi_N},
\ee
which is \Eq \eqref{raising-m} translated to the coherent-state basis. The coherent components of epsilon can be found, for example, by computing the projection of the state $\ket\epsilon = \sum_{mn} \epsilon^{(j)}_{mn}\ket{jm}\ket{jn}$ onto the basis element $\ket{j\xi}\ket{j\eta}$, or by noting that $\epsilon^{(j)}_{mn} = \sqrt{d_j}\bigl(\begin{smallmatrix} j&j&0 \\ m&n&0\end{smallmatrix}\bigr)$ and using \Eq \eqref{xi3j}. One finds
\be
\epsilon^{(j)}_{\xi\eta} = \frac{(\xi-\eta)^{2j}}{(1+|\xi|^2)^j(1+|\eta|^2)^j}
\ee
and
\be
\epsilon^{(j)\xi\eta} = \overline{\epsilon^{(j)}_{\xi\eta}}.
\ee
The integral \eqref{raising-xi} can then be evaluated using the integration formula \eqref{integration}, derived in the next section. The result is
\be
\iota\,\updown{\xi_1}{\xi_2\cdots\xi_N} = \frac{\bar\xi_1^{2j_1}P_\iota(-\bar\xi_1^{-1},\xi_2,\dots,\xi_N)}{(1+|\xi_1|^2)^{j_1}\cdots(1+|\xi_N|^2)^{j_N}}.
\ee
This can alternatively be written as
\be
\iota\,\updown{\xi_1}{\xi_2\cdots\xi_N} = \frac{P_\iota(-\bar\xi_1^{-1},\xi_2,\dots,\xi_N)}{(1+|{-}\bar\xi_1^{-1}|^2)^{j_1}\cdots(1+|\xi_N|^2)^{j_N}},
\ee
which shows that, when the intertwiner is described in terms of the vectors corresponding to the parameters $\xi$, then raising an index $\xi_a$ is equivalent to replacing the corresponding vector $\vec n_a$ with $-\vec n_a$.

\section{A polynomial representation of intertwiners}\label{intpoly}

In the previous section we found that the components of an intertwiner with respect to the coherent state basis of ${\cal H}_{j_1}\otimes\cdots\otimes{\cal H}_{j_N}$have the general form
\be
\iota_{\xi_1\cdots\xi_N} = \frac{P_\iota(\xi_1,\dots,\xi_N)}{(1+|\xi_1|^2)^{j_1}\cdots(1+|\xi_N|^2)^{j_N}}.
\ee
In this section we will show that the polynomial $P_\iota(\xi_1,\dots,\xi_N)$ can itself be regarded as a representation of the intertwiner $\iota$. In section \ref{sec:Pj} we describe a representation of $SU(2)$ in the space ${\cal P}_j$, the space of polynomials of degree $2j$ of a single complex variable. Then, in section \ref{intinPj} we show that the action of an angular momentum operator on the intertwiner $\iota$ results in an action on the polynomial $P_\iota(\xi_1,\cdots,\xi_N)$ by a differential operator representing the angular momentum operator in the space ${\cal P}_j$. From this we conclude that $P_\iota(\xi_1,\cdots,\xi_N)$, regarded as an element of ${\cal P}_{j_1}\otimes\cdots\otimes{\cal P}_{j_N}$, is the representation of the intertwiner $\ket\iota$.

\subsection{Representation of $SU(2)$ in a space of complex polynomials}\label{sec:Pj}

A convenient realization of the spin-$j$ representation of $SU(2)$ can be constructed on the space of polynomials of degree $2j$ in a complex variable $z$ (see \eg \cite{Perelomov}). We denote this space by ${\cal P}_j$; a general element of ${\cal P}_j$ has the form $f^{(j)}(z) = \sum_{k=0}^{2j} c_kz^k$. A scalar product on ${\cal P}_j$ can be defined as
\be\label{Pj prod}
\braket{f^{(j)}}{g^{(j)}} = \int d\nu_j(z)\,\overline{f^{(j)}(z)}g^{(j)}(z),
\ee
where the integration measure is
\be\label{dnu_j}
d\nu_j(z) = \frac{d_j}{\pi}\frac{d^2z}{(1+|z|^2)^{2j+2}} = \frac{d\mu_j(z)}{(1+|z|^2)^{2j}}
\ee
with $d\mu_j(z)$ the measure of \Eq \eqref{dmu_j}. A basis of ${\cal P}_j$, orthonormal under the scalar product \eqref{Pj prod}, is given by the monomials
\be\label{f_jm}
f^{(j)}_m(z) = \sqrt{\frac{(2j)!}{(j+m)!(j-m)!}}z^{j+m}
\ee
for $m=-j,-j+1,\dots,j$. An immediate consequence of the orthonormality of the functions \eqref{f_jm} is the useful identity
\be\label{integration}
\int d\nu_j(\xi)\,(1+\bar\xi z)^{2j}f^{(j)}(\xi) = f^{(j)}(z),
\ee
valid for all functions $f^{(j)}\in {\cal P}_j$.\\

A representation of $SU(2)$ on ${\cal P}_j$ is obtained by defining the action of the group element $g$ of \Eq \eqref{gsu2} on a function $f^{(j)}(z)\in {\cal P}_j$ to be
\be\label{su2pol}
D^{(j)}(g)f^{(j)}(z) = (\beta z+\bar\alpha)^{2j}f^{(j)}\biggl(\frac{\alpha z-\bar\beta}{\beta z+\bar\alpha}\biggr).
\ee
In fact, this definition extends to the group $SL(2,\C)$; for $h = \bigl(\begin{smallmatrix} a&b \\ c&d \end{smallmatrix}\bigr) \in SL(2,\C)$, the assignment
\be\label{sl2cpol}
D^{(j)}(h)f^{(j)}(z) = (bz+d)^{2j}f^{(j)}\biggl(\frac{az+c}{bz+d}\biggr)
\ee
defines a representation of $SL(2,\C)$ on ${\cal P}_j$. \\

To find the representation of the angular momentum operator on ${\cal P}_j$, we can use \Eq \eqref{sl2cpol} and the explicit matrix representation of $\vec J$ in the spin-1/2 representation (see \Eq \eqref{J1/2}) to calculate, for example
\be
e^{\epsilon J_+}f^{(j)}(z) = (1+\epsilon z)^{2j}f^{(j)}\biggl(\frac{z}{1+\epsilon z}\biggr).
\ee
Expanding both sides to first order in $\epsilon$ then reveals that
\be\label{J+}
J_+ = -z^2\frac{d}{dz} + 2jz.
\ee
By similar calculations one finds
\be\label{J0}
J_0 = z\frac{d}{dz}-j
\ee
and
\be\label{J-}
J_- = \frac{d}{dz}.
\ee
It is immediate to check that the function $f^{(j)}_m$ of \Eq \eqref{f_jm} is an eigenstate of $J_0$ with eigenvalue $m$, and hence corresponds to the state $\ket{jm}\in{\cal H}_j$ in standard physics notation. \\

For reference, let us write down the angular momentum coherent state $\ket{j\xi} = D^{(j)}(\xi)\ket{j,-{}j}$ as an element of ${\cal P}_j$. Using \Eqs \eqref{jxi} and \eqref{sl2cpol}, and noting that the state $\ket{j,-{}j}$ corresponds to the element $f^{(j)}_{-j}(z)=1$ of ${\cal P}_j$, we find
\be\label{cspoly}
f_\xi^{(j)}(z) = \frac{1}{(1+|\xi|^2)^j}(1+\bar\xi z)^{2j}.
\ee
The polynomial representing the unnormalized coherent state $|j\xi)$ of \Eq \eqref{jxi-unnorm} is simply $(1+\bar\xi z)^{2j}$. In this notation, the integration formula \eqref{integration} is expressed as
\be\label{integration2}
\int d\nu_j(\xi)\,(j\xi|jz)f^{(j)}(\xi) = f^{(j)}(z).
\ee
Let us also note that the matrix elements of the angular momentum operator between the states $|j\xi)$, computed in section \ref{spincs}, can now be written as
\be
(j\xi|J_\mu|j\eta) = J_\mu(\bar\xi)(j\xi|j\eta),
\ee
where $J_\mu(\bar\xi)$ $(\mu = +,0,-)$ denotes any of the differential operators of \Eqs \eqref{J+}--\eqref{J-} acting on the variable $\bar\xi$.

\subsection{Intertwiners as elements of ${\cal P}_j$}\label{intinPj}

We will now show how a polynomial representation of the intertwiner $\ket\iota$ can be extracted from the explicit expression of $\iota_{\xi_1\cdots\xi_N}$. The component of the intertwiner $\ket\iota$ with respect to the basis state $\ket{j_1\xi_1\otimes\cdots\otimes j_N\xi_N}$ is given by the complex conjugate of the object $\iota_{\xi_1\cdots\xi_N}$ of \Eq \eqref{iN_xi}. Therefore, $\ket\iota$ can be expanded in the basis of coherent states as
\be\label{iota-expansion}
\ket\iota = \int d\mu_{j_1}(\xi_1)\cdots d\mu_{j_N}(\xi_N)\,\overline{\iota_{\xi_1\cdots\xi_N}}\,\ket{j_1\xi_1\otimes\cdots\otimes j_N\xi_N}.
\ee
For the sake of clarity, let us first consider the action of a single angular momentum operator, acting on the first spin of the state $\ket\iota$. Projecting the resulting state on the basis element $\bra{j_1\eta_1\otimes\cdots\otimes j_N\eta_N}$, inserting the form of $\iota_{\xi_1\cdots\xi_N}$ given by \Eq \eqref{iN_xi}, and introducing the unnormalized states $|j\xi)$ from \Eq \eqref{jxi-unnorm} and the measure $d\nu_j(\xi)$ from \Eq \eqref{dnu_j}, we obtain
\begin{align}
&\bra{j_1\eta_1\otimes\cdots\otimes j_N\eta_N}J_\mu^{(1)}\ket\iota = \frac{1}{(1+|\eta_1|^2)^{j_1}\cdots(1+|\eta_N|^2)^{j_N}} \notag \\
&{}\times \int d\nu_{j_1}(\xi_1)\,\cdots d\nu_{j_N}(\xi_N)\,P_\iota(\bar\xi_1,\dots,\bar\xi_N)\,(j_1\eta_1|J_\mu|j_1\xi_1)(j_2\eta_2|j_2\xi_2)\cdots (j_N\eta_N|j_N\xi_N).
\end{align}
To perform the integral, we first recall that $(j_1\eta_1|J_\mu^{(1)}|j_1\xi_1) = J_\mu(\bar\eta_1)(j_1\eta_1|j_1\xi_1)$, and then write $(j\eta|j\xi) = (j\bar\xi|j\bar\eta)$. This allows us to evaluate the integral using \Eq \eqref{integration2}; the result is $J_\mu(\bar\eta_1)P(\bar\eta_1,\cdots,\bar\eta_N)$. We therefore conclude
\be
\bra{j_1\eta_1\otimes\cdots\otimes j_N\eta_N}J_\mu^{(1)}\ket\iota = \frac{J_\mu(\bar\eta_1)P_\iota(\bar\eta_1,\dots,\bar\eta_N)}{(1+|\eta_1|^2)^{j_1}\cdots(1+|\eta_N|^2)^{j_N}}.
\ee
By a similar calculation we can show that for any operator ${\cal O}(J^{(1)},\cdots,J^{(N)})$, constructed from the angular momentum operators by a power series expansion, we have\footnote{As indicated by \Eq \eqref{iota-expansion}, we consider an intertwiner whose all indices are lower indices. For each upper index of the intertwiner, the corresponding angular momentum operator $J^{(a)}_m$ should be replaced with $(J^{(a)}\mu)^\dagger = J^{(a)}_{-\mu}$ in ${\cal O}(J^{(1)},\cdots,J^{(N)})$ in \Eq \eqref{O(J)P}.}
\be\label{O(J)P}
\bra{j_1\xi_1\otimes\cdots\otimes j_N\xi_N}{\cal O}(J^{(1)},\cdots,J^{(N)})\ket\iota = \frac{{\cal O}\bigl(J(\bar\xi_1),\cdots,J(\bar\xi_N)\bigr)P_\iota(\bar\xi_1,\dots,\bar\xi_N)}{(1+|\xi_1|^2)^{j_1}\cdots(1+|\xi_N|^2)^{j_N}}.
\ee
That is, the action of the angular momentum operator on the state $\ket\iota$ translates to an action on the polynomial $P_\iota(\xi_1,\dots,\xi_N)$ by the differential-operator representation of the angular momentum operator on the space ${\cal P}_{j_1}\otimes\cdots\otimes{\cal P}_{j_N}$. This confirms that we can consistently interpret $P_\iota(\xi_1,\dots,\xi_N)$ as the element of ${\cal P}_{j_1}\otimes\cdots\otimes{\cal P}_{j_N}$ corresponding to the intertwiner $\ket\iota$. In this description of the intertwiner, the variable $\xi_a$ is no longer regarded as a parameter specifying an unit vector $\vec n_a$, but only as an abstract variable, whose first $2j_a$ powers span the space ${\cal P}_{j_a}$.

\section{Application to calculations in LQG}
\label{LQGapplication}

As a concerete example of applying the techniques developed in the previous sections to calculations in loop quantum gravity, let us consider the operator
\be\label{Cab}
C_{ab} = \epsilon_{ijk}\,{\rm Tr}^{(l)}\bigl(\tau_k h_{\alpha_{ab}}\bigr)J^{(a)}_i J^{(b)}_j,
\ee
which is used to construct the physical Hamiltonian in a model of loop quantum gravity deparametrized with respect to a free scalar field. (For more details, see \cite{LQGsf}.) The operator $C_{ab}$ acts on a pair of edges $e_a$ and $e_b$ sharing a node in a spin network state, the two angular momentum operators $J^{(a)}_i$ and $J^{(b)}_j$ acting on the holonomies associated with the two edges. Furthermore, $h_{\alpha_{ab}}$ denotes a holonomy around a loop $\alpha_{ab}$, which is attached to the graph of the spin network and is tangent to, but does not overlap with, the two edges $e_a$ and $e_b$. Finally, $\tau_k = -i\sigma_k/2$ denotes the anti-Hermitian generators of $SU(2)$. \\

Therefore, at the level of intertwiners described in the standard basis of magnetic indices, the action of the operator $C_{ab}$ is to replace the original intertwiner $\iota_{m_1\cdots m_N}$ at the node with a new intertwiner $(C_{ab}\iota)_{m_1\cdots m_N n_1n_2}$. The two new indices of the new intertwiner are in the end contracted with the representation matrix $D^{(l)}_{n_1n_2}(h_{\alpha_{ab}})$ of the holonomy around the loop $\alpha_{ab}$. \\

The action of the operator $C_{ab}$ on an intertwiner in the coherent-state basis can be found by evaluating the object
\be\label{C_abi_xi}
(C_{ab}\iota)_{\xi_1\cdots\xi_N\eta_1\eta_2} \equiv \braket{\epsilon_{ijk}(\tau_k^{(l)})_{\eta_1\eta_2}J^{(a)}_i J^{(b)}_j\,\iota}{j_1\xi_1\otimes\cdots\otimes j_N\xi_N},
\ee
where the indices $\eta_1$ and $\eta_2$ are eventually contracted with the indices of the representation matrix $D^{(l)}_{\eta_1\eta_2}(h_{\alpha_{ab}})$ by the integral $\int d\mu_l(\eta_1)\,d\mu_l(\eta_2)$. On grounds of \Eq \eqref{O(J)P}, we obtain from \Eq \eqref{C_abi_xi}
\be\label{Ciota}
(C_{ab}\iota)_{\xi_1\cdots\xi_N\eta_1\eta_2} = \frac{(C_{ab}P_\iota)(\xi_1,\cdots,\xi_N,\eta_1,\eta_2)}{(1+|\xi_1|^2)^{j_1}\cdots(1+|\xi_N|^2)^{j_N}(1+|\eta_1|^2)^l(1+|\eta_2|^2)^l},
\ee
where $(C_{ab}P_\iota)(\xi_1,\cdots,\xi_N,\eta_1,\eta_2)$ is a polynomial which represents the new intertwiner as an element of ${\cal P}_{j_1}\otimes\cdots\otimes{\cal P}_{j_N}\otimes{{\cal P}_l}\otimes{\cal P}_l$. An expression for $C_{ab}P_\iota$ can be derived by using \Eqs \eqref{J+}--\eqref{J-} for the angular momentum operators, and the explicit form of the matrix elements $(\tau_k^{(l)})_{\eta_1\eta_2} = -\tfrac{i}{2}\bra{l\eta_1}J_k\ket{l\eta_2}$ given by \Eqs \eqref{J+xy}--\eqref{J-xy}. Introducing the abbreviations $\partial_a = \partial/\partial\xi_a$ and $\partial^2_{ab} = \partial^2/\partial\xi_a\partial\xi_b$, we find
\begin{align}
&(C_{ab}P_\iota)(\xi_1,\cdots,\xi_N,\eta_1,\eta_2) = \frac{l}{2}(1+\eta_1\bar\eta_2)^{2l-1}\biggl[\eta_1\Bigl((\xi_a-\xi_b)\partial^2_{ab} + j_b\partial_a - j_a\partial_b\Bigr) \notag \\
&+\bar\eta_2\Bigl(-\xi_a\xi_b(\xi_a-\xi_b)\partial^2_{ab} + j_b\xi_a(\xi_a-2\xi_b)\partial_a - j_a\xi_b(\xi_b-2\xi_a)\partial_b - 2j_aj_b(\xi_a-\xi_b)\Bigr) \notag \\
&+(\eta_1\bar\eta_2-1)\Bigl(\tfrac{1}{2}(\xi_a^2-\xi_b^2)\partial^2_{ab} + j_b\xi_b\partial_a - j_a\xi_a\partial_b\Bigr)\biggr] P_\iota(\xi_1,\dots,\xi_N). \label{CP}
\end{align}
On a general level, we cannot make \Eqs \eqref{Ciota} and \eqref{CP} more explicit, since we lack a general expression for the polynomial $P_\iota(\xi_1,\dots,\xi_N)$. Let us therefore specialize to the case of computing the action of the operator $C_{12}$ on a three-valent intertwiner. In this case, the explicit form of $P_\iota(\xi_1,\xi_2,\xi_3)$ can be read off from \Eq \eqref{xi3j} as
\be
P_\iota(\xi_1,\xi_2,\xi_3) = (\xi_1-\xi_2)^{j_1+j_2-j_3}(\xi_2-\xi_3)^{j_2+j_3-j_1}(\xi_3-\xi_1)^{j_3+j_1-j_2}.
\ee
Introducing the notation $\xi_{ab} = \xi_a-\xi_b$ and $k_a = j_1+j_2+j_3-2j_a$, it is convenient to write the derivatives appearing in \Eq \eqref{CP} as
\begin{align}
\partial_1P_\iota(\xi_1,\xi_2,\xi_3) &= \biggl(\frac{k_3}{\xi_{12}} - \frac{k_2}{\xi_{31}}\biggr)P_\iota(\xi_1,\xi_2,\xi_3), \label{d1P} \\ 
\partial_2P_\iota(\xi_1,\xi_2,\xi_3) &= \biggl(\frac{k_1}{\xi_{23}} - \frac{k_3}{\xi_{12}}\biggr)P_\iota(\xi_1,\xi_2,\xi_3)
\end{align}
and
\be\label{d12P}
\partial_{12}P_\iota(\xi_1,\xi_2,\xi_3) = \biggl(-\frac{k_3^2-k^2}{\xi_{12}^2} - \frac{k_1k_2}{\xi_{23}\xi_{31}} + \frac{k_2}{k_3}{\xi_{31}\xi_{12}} + \frac{k_3k_1}{\xi_{12}\xi_{23}} \biggr)P_\iota(\xi_1,\xi_2,\xi_3).
\ee
This suggests to express the components of the intertwiner $C_{12}\iota$, given by \Eq \eqref{Ciota}, by factoring out the component of the original intertwiner, $\bigl(\begin{smallmatrix} j_1&j_2&j_3 \\ \xi_1&\xi_2&\xi_3 \end{smallmatrix}\bigr)$. We can also separate the factor $(1+\eta_1\bar\eta_2)^{2l}/(1+|\eta_1|^2)^l(1+|\eta_2|^2)^l = \braket{l\eta_2}{l\eta_1}$, which, up to a phase, is equal to $[\tfrac{1}{2}(1+\vec n_{\eta_1}\cdot\vec n_{\eta_2})]^l$. In this way, we can write the components of the new intertwiner in the form
\be\label{Ci3}
(C_{ab}\iota)_{\xi_1\xi_2\xi_3\eta_1\eta_2} = F(\xi_1,\xi_2,\xi_3,\eta_1,\eta_2)\biggl(\frac{1+\vec n_{\eta_1}\cdot\vec n_{\eta_2}}{2}\biggr)^l\begin{pmatrix} j_1&j_2&j_3 \\ \xi_1 &\xi_2 &\xi_3 \end{pmatrix},
\ee
where $\vec n_{\eta_a}$ denotes the vector corresponding to the parameter $\eta_a$, and $F(\xi_1,\xi_2,\xi_3,\eta_1,\eta_2)$ is a function which can be determined using \Eqs \eqref{CP} and \eqref{d1P}--\eqref{d12P}.\\

In the limit of large spins, the expression \eqref{Ci3} admits the following interpretation in terms of the vectors corresponding to the parameters $\xi_a$ and $\eta_a$. As discussed in section \ref{normals}, in the large-$j$ limit the 3$j$-symbol in \Eq \eqref{Ci3} is suppressed unless the vectors $\vec n_a$ corresponding to the parameters $\xi_a$ satisfy the closure condition $j_1\vec n_1+j_2\vec n_2+j_3\vec n_3=0$. Similarly, the factor $[\tfrac{1}{2}(1+\vec n_{\eta_1}\cdot\vec n_{\eta_2})]^l$ forces the two vectors $\vec n_{\eta_1}$ and $\vec n_{\eta_2}$ to be parallel to each other, when the value of $l$ is large. Assuming that the vectors $j_a\vec n_a$ close, and that $\vec n_{\eta_1} = \vec n_{\eta_2} \equiv \vec n_\eta$, we can look at the function $F$ to determine the preferred orientation of the vector $\vec n_\eta$ relative to the vectors $\vec n_a$. \\

As a first example, let us examine the case where the spins $j_1$, $j_2$ and $j_3$ are equal to a common value $j$. If we fix $\vec n_1=\tfrac{1}{2}(\sqrt 3,0,-1)$, $\vec n_2=\tfrac{1}{2}(-\sqrt 3,0,-1)$ and $\vec n_3=(0,0,1)$, we find that the absolute value of the function $F$ is given in terms of the polar angles of the vector $\vec n_\eta$ as
\be
|F(\theta_\eta,\phi_\eta)| = \sqrt{3}j^2l\Bigl(|\sin\theta_\eta\sin\phi_\eta| + {\cal O}(j^{-1})\Bigr).
\ee
This function is evidently maximized when $\theta_\eta = \pi/2$, and $\phi_\eta=\pi/2$ or $\phi_\eta=3\pi/2$, \ie when $\vec n_\eta = (0,{\pm}1,0)$. In other words, the preferred orientation of the vector $\vec n_\eta$ is orthogonal to the plane spanned by the vectors $\vec n_1$, $\vec n_2$ and $\vec n_3$. This conclusion is independent of the values of the spins $j$ and $l$, since up to subleading terms the absolute value $|F(\theta_\eta,\phi_\eta)|$ depends on $j$ and $l$ only through an overall multiplicative factor. \\

As another example we give the case $j_1=j_2=j$ and $j_3\approx \sqrt 2 j$. To satisfy the closure condition of the vectors $\vec n_a$, we now fix $\vec n_1=(0,0,1)$, $\vec n_2=(1,0,0)$ and $\vec n_3=\tfrac{1}{\sqrt 2}(-1,0,-1)$. Now the explicit expression of $|F(\theta_\eta,\phi_\eta)|$ is unenlightening, so we show a numerical plot of $|F(\theta_\eta,\phi_\eta)|/j^2l$ (which is independent of $j^2$ and $l$ at leading order) in \Fig \ref{Fplot} below. The plot indicates that the preferred orientation of the vectors $\vec n_\eta$ is still perpendicular to the plane of the vectors $\vec n_1$, $\vec n_2$ and $\vec n_3$.

\begin{figure}[tb]
	\centering
		\includegraphics[width=1\textwidth]{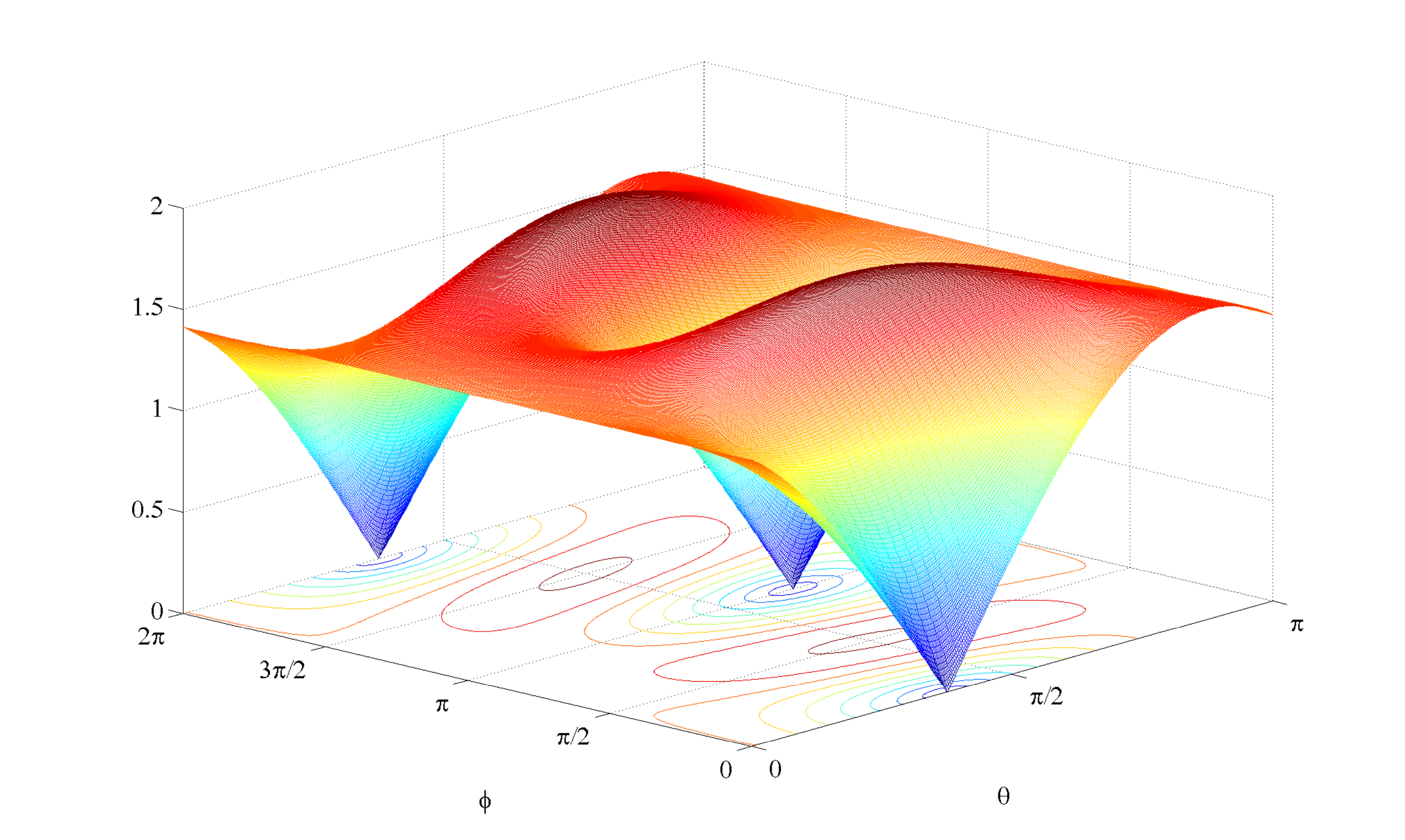}
		\caption{A plot of the function $|F(\theta_\eta,\phi_\eta)|/j^2l$, when the spins of the three-valent intertwiner $\iota$ are given by $j_1=j_2=j$ and $j_3\approx \sqrt 2j$. The function $F(\theta_\eta,\phi_\eta)$ determines the preferred orientation of the vectors $\vec n_{\eta_1} = \vec n_{\eta_2} \equiv \vec n_\eta$, created by the action of the operator $C_{12}$ on the intertwiner $\iota$. The maxima of the function $|F(\theta_\eta,\phi_\eta)|$ are located at $(\theta_\eta,\phi_\eta) = (\tfrac{1}{2}\pi,\tfrac{1}{2}\pi)$ and $(\theta_\eta,\phi_\eta) = (\tfrac{1}{2}\pi,\tfrac{3}{2}\pi)$. Both points correspond to the vector $\vec n_\eta$ orthogonal to the plane of the vectors $\vec n_1$, $\vec n_2$ and $\vec n_3$.}
	\label{Fplot}
\end{figure}

\section{Conclusions}

In this work we studied intertwiners projected onto the basis of angular momentum coherent states. The relevance of our calculations to loop quantum gravity is twofold. Firstly, the object $\iota_{\xi_1\cdots\xi_N}$, defined in \Eq \eqref{iotaxi} and given in explicit form in \Eqs \eqref{xi3j} and \eqref{xi-int4} in the case of three- and four-valent intertwiners, provides an expression for the components of the Livine--Speziale coherent intertwiner $\ket{j_1\cdots j_N;\xi_1\cdots\xi_N}$ with respect to the standard basis $\{\ket\iota\}$ of the intertwiner space. Secondly, and more importantly, by rewriting the spin network states of loop quantum gravity in terms of intertwiners projected on coherent states, operators such as the Hamiltonian can be reformulated as differential operators acting on polynomials of the complex variables $\xi_a$. \\

The primary motivation for our work is to introduce an alternative approach towards analyzing the matrix elements of the Hamiltonian operator, in which the use of $SU(2)$ recoupling theory is bypassed. By rewriting the spin network states and the Hamiltonian in the language of the complex numbers $\xi_a$, and by providing explicit expressions for the projections of intertwiners on coherent states, we make it possible to describe the action of the Hamiltonian geometrically, in terms of the vectors corresponding to the parameters $\xi_a$. This was illustrated by the example in section \ref{LQGapplication}, where we considered (the Euclidean part of) the physical Hamiltonian of a deparametrized model of loop quantum gravity \cite{LQGsf}. We computed the action of the operator on a three-valent intertwiner, and found that the preferred orientation of the two new vectors created by the Hamiltonian is orthogonal to the plane containing the three vectors of the original intertwiner. \\

Our expectation is that the techniques presented in this work will be useful for gaining a more controlled analytical understanding of the Hamiltonian in loop quantum gravity, and eventually for studying the semiclassical limit of the dynamics through an asymptotic analysis of the matrix elements of the Hamiltonian (\eg via saddle point techniques). However, the scope of the present paper is merely to show that the coherent states of angular momentum can be used to develop an alternative framework for performing calculations with the Hamiltonian. Carrying out a systematic analysis of the dynamics and obtaining a clear understanding of the geometrical content of the Hamiltonian in this framework is a question left for future work. \\

\begin{center}
 \large{\bf{Acknowledgments}}
\end{center}

The work of EA and JL was supported by the grant of Polish Narodowe Centrum Nauki nr DEC-2011/02/A/ST2/00300. EA wishes to acknowledge the John Templeton Foundation for the supporting grant \#51876. IM would like to thank the Jenny and Antti Wihuri Foundation for support.

\appendix

\section{Selected facts about $SU(2)$}\label{app:su2}

In this appendix we recall some results from $SU(2)$ representation theory, which will be needed in the main text. 

\subsection{Parametrizations of $SU(2)$}

A general element $g\in SU(2)$ has the form
\be\label{gsu2}
g = \begin{pmatrix} \alpha & \beta \\ -\bar\beta & \bar\alpha \end{pmatrix} \qquad \text{where} \quad |\alpha|^2 + |\beta|^2 = 1.
\ee
The group element can be parametrized in terms of an angle $\chi$ and an axis of rotation $\vec n$ as
\be
g = e^{-i\chi\vec n\cdot\vec J} = \cos\frac{\chi}{2} - i\sin\frac{\chi}{2}(\vec n\cdot\vec\sigma),
\ee
where $\vec J = \vec\sigma/2$ is the angular momentum operator for $j=1/2$, and $\vec\sigma = (\sigma_x,\sigma_y,\sigma_z)$ are the Pauli matrices. \\

Further parametrizations of the group element can be obtained using the operators $J_\pm = J_x\pm iJ_y$ and $J_0=J_z$. One has the decompositions
\be\label{-0+dec}
g = e^{\lambda J_-}e^{\mu J_0}e^{\nu J_+}
\ee
and
\be\label{+0-dec}
g = e^{\xi J_+}e^{\eta J_0}e^{\zeta J_-}
\ee
where the parameters are given by
\be
e^{\mu/2} = \alpha \qquad \lambda = -\frac{\bar\beta}{\alpha} \qquad \nu = \frac{\beta}{\alpha}
\ee
and
\be
e^{-\eta/2} = \bar\alpha \qquad \xi = \frac{\beta}{\bar\alpha} \qquad\zeta = -\frac{\bar\beta}{\bar\alpha}.
\ee
The relations \eqref{-0+dec} and \eqref{+0-dec} can be verified by direct calculation in the fundamental representation of the group, using an explicit matrix representation for the angular momentum operators\footnote{The standard representation of $\vec J$ is given in the basis in which $J_z$ is diagonal, and reads 
\be\label{J1/2}
J_x = \frac{1}{2}\begin{pmatrix} 0&1 \\ 1&0 \end{pmatrix}, \qquad
J_y = \frac{1}{2}\begin{pmatrix} 0&-i \\ i&0 \end{pmatrix}, \qquad
J_z = \frac{1}{2}\begin{pmatrix} 1&0 \\ 0&-1 \end{pmatrix}.
\ee }. Note that in general, none of the matrices on the right-hand side of \Eqs \eqref{-0+dec} and \eqref{+0-dec} are elements of $SU(2)$, though they do belong $SL(2,\C)$.

\subsection{The spin-$j$ representation as a symmetric tensor product}\label{su2prod}

The fundamental (spin-1/2) representation of $SU(2)$ is realized by letting the matrix $g = g^A_{\;\; B}$ of \Eq \eqref{gsu2} act on the space ${\cal H}_{1/2} \cong \C^2$ spanned by the vectors $v^A$, the index $A$ taking the two values $A=+,-$. A realization of the spin-$j$ representation that is sometimes useful is obtained by considering the space ${\cal H}_j$ as the completely symmetric part of the tensor product of $2j$ copies of the space ${\cal H}_{1/2}$. Hence, ${\cal H}_j$ is spanned by the $d_j = 2j+1$ vectors $v^{(A_1\cdots A_{2j})} \equiv v^m$; the correspondence between the completely symmetric index $(A_1\cdots A_{2j})$ and the magnetic index $m$ typically used in the physics literature is given by $m = \half\sum_{i=1}^{2j} A_i$. The representation of an element $g\in SU(2)$ on the space ${\cal H}$ is then given by
\be\label{Djprod}
D^{(j)}_{mn}(g) = D^{(j)}_{(A_1\cdots A_{2j})(B_1\cdots B_{2j})}(g) = g^{A_1}_{\quad (B_1}\cdots g^{A_{2j}}_{\quad B_{2j})}.
\ee
Intertwiners -- \ie invariant tensors on ${\cal H}_{j_1}\otimes\cdots\otimes{\cal H}_{j_N}$ -- are constructed using the tensor $\epsilon_{AB}$, which is the only invariant tensor in the fundamental representation\footnote{Together with $\delta^A_B$ and $\epsilon^{AB}$, which are obtained by raising one or both indices of $\epsilon_{AB}$. Indices are raised and lowered according to the conventions
\be
v^A = \epsilon^{AB}v_B, \qquad v_A = v^B\epsilon_{BA}. 
\ee}. For example, the three-valent intertwiner between representations $j_1$, $j_2$ and $j_3$ is given by
\begin{align}
\iota^{(j_1j_2j_3)}_{m_1m_2m_3} &= \iota_{(A_1\cdots A_{2j_1})(B_1\cdots B_{2j_2})(C_1\cdots C_{2j_3})} \notag \\
&= N_{j_1j_2j_3}\epsilon_{A_1B_1}\cdots\epsilon_{A_aB_a}\epsilon_{B_{a+1}C_1}\cdots\epsilon_{B_{a+b}C_b}\epsilon_{C_{b+1}A_{a+1}}\cdots\epsilon_{C_{b+c}A_{a+c}}, \label{iota3}
\end{align}
where a complete symmetrization of the indices $(A_1\cdots A_{2j_1})$, $(B_1\cdots B_{2j_2})$ and $(C_1\cdots C_{2j_3})$ is implied on the right-hand side. The numbers $a$, $b$ and $c$ are given by
\be
a = j_1+j_2-j_3, \qquad b = j_2+j_3-j_1, \qquad c = j_3+j_1-j_2,
\ee
and the normalization factor is
\be\label{Njjj}
N_{j_1j_2j_3} = \sqrt{\frac{(2j_1)!(2j_2)!(2j_3)!}{(j_1+j_2+j_3+1)!(j_1+j_2-j_3)!(j_2+j_3-j_1)!(j_3+j_1-j_2)!}}.
\ee
Intertwiners of higher valence can be constructed by contraction of three-valent intertwiners. For example, an orthonormal basis in the four-valent intertwiner space between representations $j_1$, $j_2$ $j_3$ and $j_4$ is given by the objects
\be\label{iota4}
\iota^{(k)}_{(A_1\cdots A_{2j_1})\cdots(D_1\cdots D_{2j_4})} = \sqrt{d_k}\iota_{(A_1\cdots A_{2j_1})(B_1\cdots B_{2j_2})(E_1\cdots E_{2k})}\iota\updown{(E_1\cdots E_{2k})}{(C_1\cdots C_{2j_3})(D_1\cdots D_{2j_4})}.
\ee
The three-valent intertwiner \eqref{iota3} can be represented graphically by introducing a notation in which 
\begin{itemize}
\item $\epsilon_{AB}$ or $\epsilon^{AB}$ is represented by a line with an arrow pointing from index $A$ to $B$;
\item $\delta^A_B = \epsilon^{AC}\epsilon_{BC}$ is represented by a line with no arrow;
\item Contraction of an index is represented by connecting two lines at the contracted index.
\end{itemize}
Then we can write
\be\label{i3graphic}
\iota_{(A_1\cdots A_{2j_1})(B_1\cdots B_{2j_2})(C_1\cdots C_{2j_3})} = N_{j_1j_2j_3}\;\makeSymbol{\includegraphics[scale=0.75]{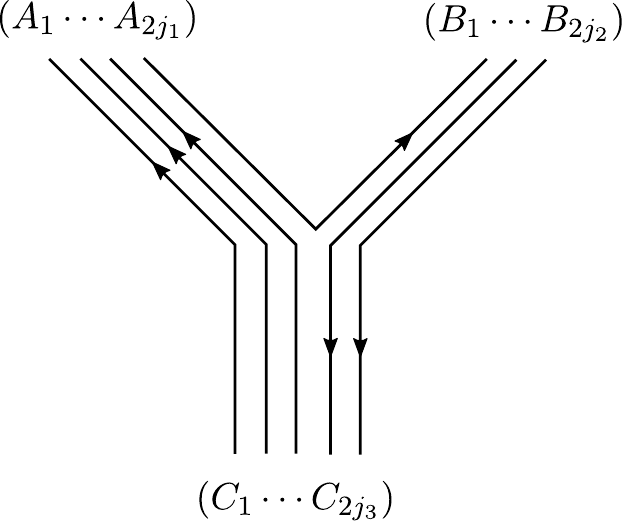}}
\ee
where each external group of lines is understood to be completely symmetrized. In this notation, the four-valent intertwiner \eqref{iota4} is given by
\be\label{i4graphic}
\iota^{(k)}_{(A_1\cdots A_{2j_1})\cdots(D_1\cdots D_{2j_4})} = \sqrt{d_k}N_{j_1j_2k}N_{kj_3j_4}\;\makeSymbol{\includegraphics[scale=0.75]{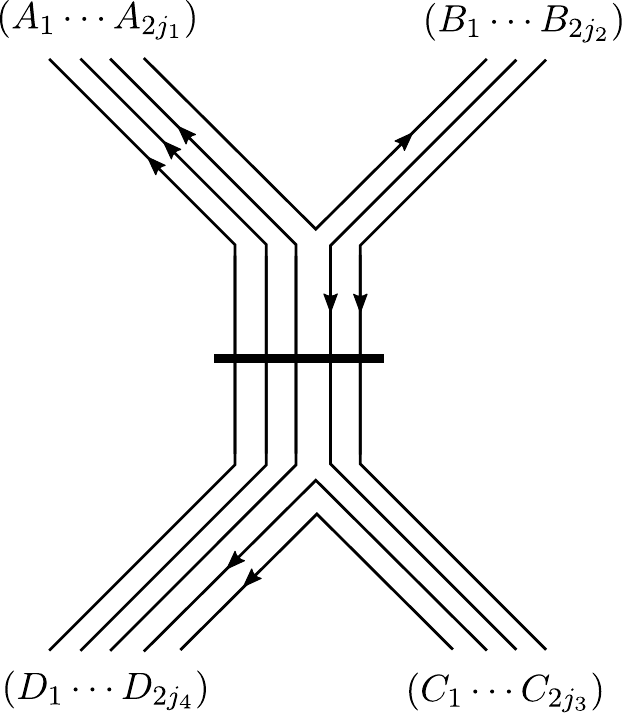}}
\ee
where we draw the vertical bar to indicate explicitly a complete symmetrization of the internal group of lines.

\begin{lrbox}{\mybox}
$\begin{pmatrix} j_1&j_2&j_3 \\ \xi_1&\xi_2&\xi_3 \end{pmatrix}$
\end{lrbox}

\section{A polynomial calculation of \usebox{\mybox}}\label{app:poly}

The complex polynomial representation of $SU(2)$, described in section \ref{sec:Pj}, provides an alternative way of calculating the components of the 3$j$-symbol in the basis of coherent states. In this appendix we will perform the calculation, following a similar calculation carried out in \cite{BasuKar} for the Clebsch--Gordan coefficient. In this way we provide both another example of using the polynomial formalism, and an alternative derivation of \Eq \eqref{xi3j}. \\

As defined in \Eq \eqref{iDDD}, the coherent components of the 3$j$-symbol are given by
\be
\begin{pmatrix} j_1&j_2&j_3 \\ \xi_1&\xi_2&\xi_3 \end{pmatrix} = \begin{pmatrix} j_1&j_2&j_3 \\ m_1&m_2&m_3 \end{pmatrix}D^{(j_1)}_{m_1,-{}j_1}(\xi_1)D^{(j_2)}_{m_2,-{}j_2}(\xi_2)D^{(j_3)}_{m_3,-{}j_3}(\xi_3).
\ee
Using the $SU(2)$ invariance of the standard 3$j$-symbol, this can be rewritten as
\be\label{coh3j-2}
\begin{pmatrix} j_1&j_2&j_3 \\ \xi_1&\xi_2&\xi_3 \end{pmatrix} = \sum_{m_1m_2} \begin{pmatrix} j_1&j_2&j_3 \\ m_1&m_2&-j_3 \end{pmatrix}D^{(j_1)}_{m_1,-{}j_1}(\xi_3^{-1}\xi_1)D^{(j_2)}_{m_2,-{}j_2}(\xi_3^{-1}\xi_2).
\ee
Computing the products of two coherent state rotations by means of \Eqs \eqref{g-1} and \eqref{g1g2}, and defining the state
\be\label{Phi_0}
\ket{\Phi_0} = \sum_{m_1m_2} \begin{pmatrix} j_1&j_2&j_3 \\ m_1&m_2&-j_3 \end{pmatrix} \ket{j_1m_1}\ket{j_2m_2},
\ee
we can write \Eq \eqref{coh3j-2} as
\be\label{prod2}
\begin{pmatrix} j_1&j_2&j_3 \\ \xi_1&\xi_2&\xi_3 \end{pmatrix} = e^{-ij_1\beta_{13}-ij_2\beta_{23}}\braket{\Phi_0}{j_1\xi_{13}\otimes j_2\xi_{23}}
\ee
where
\be\label{parameters}
\xi_{13} = \frac{\xi_1-\xi_3}{1+\xi_1\bar\xi_3}, \qquad e^{i\beta_{13}/2} = \frac{1+\bar\xi_1\xi_3}{|1+\bar\xi_1\xi_3|}
\ee
(and the same for $\xi_{23}$ and $\beta_{23}$). Transforming the 3$j$-symbol in \Eq \eqref{Phi_0} into a Clebsch--Gordan coefficient, we see that
\be
\ket{\Phi_0} = \frac{(-1)^{j_1-j_2+j_3}}{\sqrt{d_{j_3}}}\sum_{m_1m_2} C^{j_1j_2j_3}_{m_1m_2j_3} \ket{j_1m_1}\ket{j_2m_2} = \frac{(-1)^{j_1-j_2+j_3}}{\sqrt{d_{j_3}}}\ket{j_1j_2;j_3j_3}
\ee
\ie up to a numerical factor, $\ket{\Phi_0}$ is the state of total spin $j_3$ and maximal magnetic number $m_3=j_3$, constructed from spins $j_1$ and $j_2$. \\

Our aim now is to compute the scalar product $\braket{\Phi_0}{j_1\xi_{13}\otimes j_2\xi_{23}}$ in \Eq \eqref{prod2} using the polynomial representation of section \ref{sec:Pj}, considering $\ket{\Phi_0}$ and $\ket{j_1\xi_{13}\otimes j_2\xi_{23}}$ as polynomials in the space ${\cal P}_{j_1}\otimes{\cal P}_{j_2}$. The coherent states are given in polynomial form by \Eq \eqref{cspoly}. To find the polynomial representation of the state $\ket{\Phi_0}$, we write the defining equations of the state $\ket{j_1j_2,j_3j_3}$, $J_+\ket{j_1j_2,j_3j_3}=0$ and $J_0\ket{j_1j_2,j_3j_3}=j_3\ket{j_1j_2,j_3j_3}$, as differential equations for a function $f_{j_3}(z_1,z_2)$ in ${\cal P}_{j_1}\otimes{\cal P}_{j_2}$, using the representation of the angular momentum operators from \Eqs \eqref{J+}--\eqref{J-}. The equations are
\begin{align}
&\biggl(-z_1^2\frac{\partial}{\partial z_1} + 2j_1z_1 - z_2^2\frac{\partial}{\partial z_2} + 2j_2z_2\biggr)f_{j_3}(z_1,z_2) = 0 \\
&\biggl(z_1\frac{\partial}{\partial z_1} - j_1 + z_2\frac{\partial}{\partial z_2} - j_2\biggr)f_{j_3}(z_1,z_2) = j_3f_{j_3}(z_1,z_2).
\end{align}
The solution which is an element of ${\cal P}_{j_1}\otimes{\cal P}_{j_2}$ is given by
\be
f_{j_3}(z_1,z_2) = M_{j_1j_2j_3}z_1^{j_1-j_2+j_3}z_2^{j_2-j_1+j_3}(z_1-z_2)^{j_1+j_2-j_3}.
\ee
The correct value of the normalization constant is $M_{j_1j_2j_3} = \sqrt{d_{j_3}}N_{j_1j_2j_3}$, with $N_{j_1j_2j_3}$ given by \Eq \eqref{Njjj}. (The phase of $M_{j_1j_2j_3}$ is chosen according to the Condon--Shortley convention, in which the Clebsch--Gordan coefficient $C^{j_1j_2j_3}_{j_1,j_3{-}j_1,j_3}$ -- equivalently, the coefficient of $z_1^{2j_1}z_2^{j_2+j_3-j_1}$ in the expansion of the function $f_{j_3}(z_1,z_2)$ -- is real and positive.) \\

It follows that the representation of the state $\ket{\Phi_0}$ as a polynomial is given by
\be
\phi_0(z_1,z_2) = N_{j_1j_2j_3}(-z_1)^{j_1-j_2+j_3}z_2^{j_2-j_1+j_3}(z_1-z_2)^{j_1+j_2-j_3}.
\ee
The scalar product in \Eq \eqref{prod2} can therefore be computed as
\begin{align}
&\braket{\Phi_0}{j_1\xi_{13}\otimes j_2\xi_{23}} = \int d\nu_{j_1}(z_1)\,d\nu_{j_2}(z_2)\,\overline{\phi_0(z_1,z_2)}f^{(j_1)}_{\xi_{13}}(z_1)f^{(j_2)}_{\xi_{23}}(z_2) \notag \\
&= \frac{N_{j_1j_2j_3}}{(1+|\xi_{13}|^2)^{j_1}(1+|\xi_{23}|^2)^{j_2}} \int d\nu_{j_1}(z_1)\,d\nu_{j_2}(z_2)\,(1+\xi_{13}z_1)^{2j_1}(1+\xi_{23}z_2)^{2j_2} \notag \\
&\hspace{7cm}{}\times(-\bar z_1)^{j_1-j_2+j_3}\bar z_2^{j_2-j_1+j_3}(\bar z_1-\bar z_2)^{j_1+j_2-j_3}.
\end{align}
In the integral, the function multiplying $(1+\xi_{13}z_1)^{2j_1}(1+\xi_{23}z_2)^{2j_2}$ is a polynomial of order $2j_a$ in each $\bar z_a$. Hence, after changing the variables of integration from $z_1$ and $z_2$ to $\bar z_1$ and $\bar z_2$, the integrals can be immediately evaluated using \Eq \eqref{integration}. This yields
\be
\braket{\Phi_0}{j_1\xi_{13}\otimes j_2\xi_{23}} = \frac{N_{j_1j_2j_3}}{(1+|\xi_{13}|^2)^{j_1}(1+|\xi_{23}|^2)^{j_2}}(-\xi_{13})^{j_1-j_2+j_3}\xi_{23}^{j_2-j_1+j_3}(\xi_{13}-\xi_{23})^{j_1+j_2-j_3}.
\ee
When this is inserted into \Eq \eqref{prod2}, and the result expressed in terms of $\xi_1$, $\xi_2$ and $\xi_3$ using \Eq \eqref{parameters}, we recover \Eq \eqref{xi3j}.

\renewcommand{\bibname}{References}

\end{document}